# Designing sustainable high-entropy alloys directly from mixed oxides: A thermodynamic perspective

Albert Linda*[,1], Shiv Shankar*[,1], Dierk Raabe

Max Planck Institute for Sustainable Materials, Max-Planck-Str. 1, 40237, Düsseldorf, Germany

*Corresponding author: a.linda@mpi-susmat.de; s.shankar@mpi-susmat.de

[1] These authors contributed equally to this work

## Abstract

Hydrogen-based direct reduction (HyDR) of mixed oxide precursors offers an alternative route to produce sustainable alloys, bypassing the sequential extraction, refining, and remelting steps that dominate the carbon footprint of conventional metallurgy and individual metal production. However, the thermodynamic limits and most effective $H_2/H_2O$ ratio needed for full metallization of multicomponent oxides are not yet established. This is because the formation of metallic and oxide solid solutions alters the chemical activity of both the reduced metallic phases and the unreduced oxides, coupling their equilibria in ways that cannot be deduced from the stability of the pure individual oxides. Herein, we develop a CALPHAD-based thermodynamic framework to establish the composition–temperature–$H_2/H_2O$ window for complete metallization of CoCrFeMnNi-based oxide mixtures. Mn and Cr generally increase oxide stability, while Co, Fe, and Ni decrease it, allowing reduction at lower $H_2/H_2O$ ratios. The effect of each element is not fixed; it depends on the starting mixed oxide composition and can reverse when the last oxide to reduce changes. Furthermore, we show that the reduction pathway complexity does not correlate with the difficulty of removing the final oxygen: a system with few intermediate transformation steps can still terminate as a highly stable residual oxide, whereas extensive partial reduction may leave the stability of the final residual oxide essentially unchanged. By linking residual-oxide stability to the $H_2/H_2O$ ratio, the framework identifies the thermodynamic limits and provides guidelines for designing sustainable multicomponent alloys directly from mixed oxide precursors.

## 1 Introduction

The transition towards sustainable metal production requires a paradigm shift in current metallurgical processing routes [1–3]. The metallurgical sector currently accounts for ~40% of global industrial greenhouse gas emissions, largely due to its continued reliance on fossil-based reductants [4–6]. Hydrogen-based direct reduction (HyDR) of mixed oxides offers a promising alternative for sustainable alloy synthesis, especially when the hydrogen is produced from renewable sources [7–9]. Unlike fossil reductants, hydrogen produces water ($H_2O$) instead of carbon dioxide ($CO_2$) as the redox product [10–13]. Beyond eliminating these direct $CO_2$ emissions, HyDR of mixed oxides combines reduction and alloying in a one-step process, reducing energy use compared to conventional multi-step synthesis routes [9,14,15].

Mixed oxide reduction has attracted significant attention since the early work of Barshchevskaya and Radomysel'skii on the HyDR of Fe-Cr oxides [7,16]. They proposed that mixing a stable oxide with a less stable one or with an alloying element can lower its reduction temperature by more than 200 °C. This concept has since motivated studies of binary, ternary, and multicomponent oxide systems. Herbel *et al*. reduced Cr-Ni oxide mixtures with hydrogen, carbon, or both, and found that Ni promotes $Cr_2O_3$ reduction because the newly formed Cr dissolves in Ni, lowering its chemical activity in solid solution [17]. Al-Buainain and Dunand reduced Cr-Fe-Ni oxides to make Fe-18Cr-8Ni (wt.%) and Fe-18Cr-35Ni (wt.%) stainless steels achieving full metallization at 1300 °C. Extending this approach to a multicomponent system, Kenel *et al*. reduced 3D-extruded inks of Co-Cr-Fe-Ni oxide mixtures in pure hydrogen and obtained a metallic alloy with a 99.6 ± 0.1% relative density at 1100 °C [18].

More recently, Shankar *et al*. developed a thermodynamic framework to predict the reducibility of Fe–Cr oxide mixtures as a function of composition and reduction temperature [19]. However, the framework did not account for the external thermodynamic driving force, i.e., the $H_2/H_2O$ ratio, which is required to predict whether reduction is thermodynamically feasible. Furthermore, in-situ HyDR of Co–Fe–Mn–Ni oxides revealed that reduction proceeds through complex phase-evolution pathways, in which transient SPINEL and HALITE phases form before forming body-centered cubic (BCC) and face-centered cubic (FCC) metallic phases [20,21]. Beyond synthetic precursors, naturally occurring ores and artificial minerals (urban-mine minerals extracted and enriched from waste streams) are also suited yet complex feedstock types containing multiple cations[6]. Knowing how these cations interact and shift the final equilibrium is essential for designing sustainable alloys from such feedstocks.

Therefore, mixed-oxide reduction is not a simple superposition of the reduction reactions of the individual oxides. Instead, it is a coupled phase transformation process governed by cation redistribution, changes in chemical activity, elemental partitioning and the evolving thermodynamic stability of oxide and metal phases. A predictive model must therefore do more than check if a single oxide can be reduced. It must quantify (i) how the addition of each component affects the thermodynamic stability of oxide mixtures, (ii) how stabilizing and destabilizing interactions evolve with composition, (iii) the extent of solid-solution formation, (iv) the energetics of the intermediate reduction steps and, where they can be resolved, their sequence, and (v) the reduction potential ($H_2/H_2O$ ratio) required for achieving complete metallization. These considerations motivated us to develop a unified thermodynamic framework to predict the reducibility of multicomponent oxide mixtures and identify their thermodynamic reduction windows.

The Co–Cr–Fe–Mn–Ni composition space contains the principal elements of several technologically important alloy classes, including CoCrFeMnNi-based high-entropy alloys [22,23] stainless steels [24], medium-Mn advance high-strength steels [25] and Fe–Ni–Co-based Invar alloys [26]. This quinary system is a particularly stringent test for such a framework because it combines several competing oxide chemistries, ranging from readily reducible oxides (FeO, $Co_3O_4$, NiO) to very stable ones ($Cr_2O_3$, MnO). Mn usually stabilizes HALITE-type oxides, Cr strongly prefers CORUNDUM-type structures, and Co, Fe, Mn and Ni can form multicomponent SPINEL and HALITE solid solutions [20,21,27,28]. The resulting competition between oxide stability, cation redistribution and metallization cause the controlling (last-to-reduce) oxide to vary with composition and temperature. Experiments have so far mapped out phases and pathways for a few specific compositions, but cannot cover the reduction landscape across the full quinary space. This is because reduction depends simultaneously on oxide composition, temperature, precursor state, $H_2/H_2O$ ratio, solid-solution formation and the driving force of successive reduction steps. Mapping all these variables experimentally would need an impractically large number of compositions and processing conditions.

Herein, we propose a thermodynamic framework based on CALPHAD (CALculation of PHAse Diagrams) approach to map this multidimensional design space. The framework links oxide composition and reduction conditions, with predictions checked against published data. By treating oxide composition as a design variable rather than simply as a precursor constraint, the framework identifies oxide mixtures that are easier to reduce and guides the choice of composition and reduction conditions for low-temperature alloy formation. More broadly, this approach establishes a general thermodynamic strategy for one-step oxide-to-alloy design, in which precursor composition and reducing environment are co-designed to directly produce the targeted alloy from mixed oxide.

# 2 Computational methodology

## 2.1 Thermodynamic model phase selection and composition sampling

We conducted the thermodynamic calculations with Thermo-Calc 2025a software [29] using the Metal Oxide Solutions (TCOX10) database for the oxide phases. For each composition and temperature, we assumed a single homogeneous FCC solid solution. This matches the equilibrium structure of the Cantor alloy (equiatomic Co–Cr–Fe–Mn–Ni) at temperatures above ~700 °C [23,30]. The FCC structure is therefore an imposed modelling assumption for all other compositions, not a calculated result; its validity range is discussed later. The calculations were automated through TC-Python, the Python interface to Thermo-Calc.

The FCC restriction defines a conditional equilibrium. Observation of FCC stability near the (equimolar) Cantor composition does not guarantee it across the whole composition range, especially near Fe, Cr, or Mn rich limits or below 900 °C. Competing BCC, ordered, or liquid metal phases could change activities and the oxide boundary. Accordingly, the present thresholds should be interpreted within the specified metallic-phase model; unrestricted calculations would be needed as an independent sensitivity testbed and before applying them to other alloys with different equilibrium structure. The limitation is supported by long-duration annealing experiments: Otto et al. observed decomposition of initially single-phase CrMnFeCoNi after 500 days at 500 °C and 700 °C, whereas it remained single-phase at 900 °C [31]. Thus, the FCC prototype is an appropriate reference state, but its persistence cannot be assumed to establish global equilibrium throughout the present temperature and composition range.

Database coverage represents a second, independent limitation. Chen et al. noted that metastable phases and extrapolation from lower-order data are key challenges in CALPHAD databases for high-entropy alloys [32]. Although their analysis focused on metallic databases and does not directly assess the validity of TCOX10, the underlying assessment principle is equally relevant here: dense sampling within a database improves the resolution of its predictions without independently establishing their accuracy in poorly assessed multicomponent regions.

No oxide phase was explicitly pre-selected. Instead, the CALPHAD global Gibbs free energy minimization algorithm simultaneously evaluates the permitted phases of the selected database subject to the metallic-phase restriction and determines the phase assemblage (the set of coexisting phases) that minimizes the total Gibbs energy at the specified composition, temperature, and oxygen chemical potential, μ(O), the partial molar Gibbs free energy of oxygen. The TCOX10 database contains Gibbs energy functions and sublattice-model parameters for the following oxide solution phases relevant to the Co–Cr–Fe–Mn–Ni–O system:

An oxidation-onset event is one composition–temperature calculation in which the first oxide becomes stable as the imposed oxygen chemical potential is increased from the metallic side. Note that this is not necessarily the experimentally observed start of oxidation. In the reverse equilibrium direction, under decreasing oxygen chemical potential conditions, the same metal/oxide boundary is the final boundary that must be crossed to achieve the complete reduction of the residual oxide. The phase counts below refer to the 35,934 valid Fe-bearing onset records (Fe > 0 (at.%, calculated on the metal-only composition basis)) above 500–1500 °C; the full dataset also includes Fe-free compositions. For conciseness, each oxide mixture is denoted by its metal constituents: Co–Fe–Mn denotes a mixed oxide whose cations are Co, Fe and Mn, and a composition such as $Co_{10}Fe_{50}Mn_{40}$ gives the cation fractions (at.%) on the metal-only basis. The same label therefore refers to both the oxide precursor and the final alloy after complete reduction. Here, the frequency of an oxide phase is the number of onset records (composition–temperature points of the sampling grid) in which that phase is the first oxide, divided by the total number of valid onset records; it is not a phase amount within a specimen.

A phase boundary is a change between adjacent calculated equilibrium phase assemblages at fixed metal composition and temperature as oxygen chemical potential is varied. The boundary index is the integer identifier assigned to a detected boundary in the stored scan: index 1 denotes the first metal-to-oxide boundary encountered from the metallic side, and higher indices identify subsequent stored boundaries. An index of 0 means no boundary was found, not a physical transition. The index is neither a phase fraction nor elapsed reaction time; higher-index energy ordering must be verified before assigning a reduction sequence. Here, the onset phase is the oxide recorded on the oxidized side of index 1, and the final reduction bottleneck is the residual-oxide boundary that must be crossed to reach the modeled metallic state.

Uppercase names such as HALITE, SPINEL and CORUNDUM are database phase labels. In the structural descriptions used below, ‘tet’ and ‘oct’ denote tetrahedrally and octahedrally coordinated cation sites, respectively; a sublattice is a set of equivalent crystallographic sites. A monoxide has a 1:1 metal-to-oxygen ratio; a sesquioxide has a 2:3 ratio. A phase field is the range of conditions over which a phase or phase assemblage is stable.

**HALITE** (rock-salt, B1 structure, $Fm\bar{3}m$): A multi-component monoxide solid solution $(^{oct}[Co^{2+}, Fe^{2+}, Mn^{2+}, Ni^{2+}])_1(O^{2-})_1$. The cation occupancies are determined by minimizing the Gibbs energy at each $(T,\mu(O))$ point. In Mn-rich alloys this phase is dominated by MnO; in Fe-rich, Mn-lean compositions by FeO. HALITE is the most frequently encountered onset phase in the dataset, comprising 42.4% of valid Fe-bearing onset events (n = 15,225).

**SPINEL** ($AB_2O_4$, $Fd\bar{3}m$): A two-sublattice solid solution $(^{tet}[Co, Fe, Mn, Ni])_1(^{oct}[Co, Cr, Fe, Mn])_2O_4$, covering the full family of transition-metal SPINELS including $CoFe_2O_4$, $FeMn_2O_4$, $MnFe_2O_4$, $MnCr_2O_4$, and $FeCr_2O_4$. Normal and inverse configurations are captured through the two-sublattice formalism. SPINEL accounts for 27.6% of onset events (n = 9,908) over the calculated range of 500–1500 °C. Within the valid Fe-bearing population, its frequency is 41.1% at 500 °C, 32.9% at 700 °C, 25.3% at 900 °C and 21.1% at 1100 °C. SPINEL therefore becomes less frequent over 500–1100 °C; its persistence at a particular composition must be read from the corresponding ternary phase map.

**CORUNDUM** (rhombohedral, $R\bar{3}c$): A sesquioxide solid solution $(Cr, Fe)_2O_3$, covering the complete range between $Cr_2O_3$ (eskolaite) and $Fe_2O_3$ (hematite), which form a continuous solid solution. CORUNDUM is the onset phase in Cr-bearing, Mn-free compositions and accounts for 28.6% of onset events (n = 10,260).

**CUB_A13** (cubic bixbyite, C-type $M_2O_3$, $Ia\bar{3}$): The high-temperature cubic polymorph of $Mn_2O_3$, encountered exclusively in Mn-dominant ($x_{Mn} \gtrsim 80$ at.%) compositions over 500–1150 °C in the valid Fe-bearing dataset. It accounts for 1.5% of onset events (n = 540) and frequently co-exists with HALITE at the onset boundary (CUB_A13|HALITE, n = 540 coexisting assemblages).

**Minor phases** (combined < 0.1%): $M_2O_3C$ (cubic C-type sesquioxide), $NiMnO_3$ (ilmenite-type), and $Ni_6MnO_8$-type phases. At high temperatures the database also includes an IONIC_LIQ oxide melt phase, which can appear as the stable oxide beyond ~ 1400°C for certain compositions.

Each solution phase is described by a sublattice model [33] in which the Gibbs free energy of mixing includes both the ideal configurational entropy of the sublattice occupancies and excess interaction parameters fitted to experimental phase equilibria. The key consequence for mixed oxide ΔG estimation is that the computed onset ΔG is not the formation energy of a fixed stoichiometric compound but the Gibbs energy at the stability boundary of a composition-optimized, multi-component solid solution: the cation distribution within the onset phase (e.g., the Mn:Fe ratio in HALITE, or the normal/inverse degree in SPINEL) adjusts automatically to the thermodynamically optimal value at each ($T$,$\mu$(O)) point. The compound energy formalism (CEF) supplies a thermodynamic representation of crystallographic sublattices; the thermodynamic literature on complex oxides also cautions against attributing phase stability to configurational entropy alone [33,34].

The size of the resulting calculation dataset, the screening of phase labels and the coverage of the composition–temperature pairs are summarized in the Supporting Information (section 'Calculation dataset and coverage').

The Co–Cr–Fe–Mn–Ni simplex was sampled with a non-uniform grid: each axis was discretized in 5 at.% steps below 10 at. % and in 10 at. % steps above, giving values {0, 5, 10, 20, ..., 100} at.%. All combinations ( $x_{Co}, x_{Cr}, x_{Fe}, x_{Mn}$ ) satisfying $\sum x_i \leq 100$ were enumerated, with $x_{Ni} = 100 - \sum_{i \neq Ni} x_i$, yielding 2,248 distinct alloy compositions from the primary quinary grid; including additional binary and ternary sub-system compositions computed with finer sub-system grids and gap-fill calculations, the full dataset spans 3,526 unique compositions. Finer spacing below 10 at. % captures the known sensitivity of oxide onset to minority additions of reactive elements such as Mn and Cr [35]. Calculations were performed at twenty-one temperatures in 50 °C steps from 500 °C to 1500 °C, with incomplete composition–temperature coverage in the assembled checkpoint.

## 2.2 Boundary extraction gas conversion and uncertainty

Oxide-onset thermodynamics were extracted using the oxygen chemical-potential scanning strategy introduced by Shang *et al.* [36]. At each fixed alloy composition and temperature *T*, the oxygen chemical potential $\mu$(O) was swept from -600 to -100 kJ mol$^{-1}$ O in up to 200 coarse steps, with a refining, adaptive 25-step refinement window around each detected phase-boundary crossing. The coarse step is 2.5 kJ mol$^{-1}$ O (500 kJ mol$^{-1}$ O divided into 200 steps), so a boundary that falls between two coarse steps is located to within ±1.25 kJ mol$^{-1}$ O (±2.5 kJ mol$^{-1}$ $O_2$). Regions narrower than 1% of the scan range (5 kJ mol$^{-1}$ O) are rescanned in 25 steps over a window extended by four coarse steps on each side (at most 25 kJ mol$^{-1}$ O), which refines the resolution to about 1 kJ mol$^{-1}$ O (±0.5 kJ mol$^{-1}$ O, i.e., ±1 kJ mol$^{-1}$ $O_2$). These uncertainties are small compared with the composition-dependent differences in onset energy of tens to hundreds of kJ mol$^{-1}$ $O_2$ discussed below; at 1100 °C, ±2.5 kJ mol$^{-1}$ $O_2$ changes the required $H_2/H_2O$ ratio by a factor of about 1.1. A boundary is detected when the stable equilibrium phase assemblage changes between consecutive $\mu$(O) steps; the onset (boundary index = 1) is the first metal-to-oxide transition encountered as $\mu$(O) increases from the metallic side. The onset boundary is characterized by the oxygen activity returned by the equilibrium calculation.

$$\Delta G_{onset} = \frac{RT}{1000}\ln\left(\frac{p_{O_2}}{p^\circ}\right) = \frac{2\Delta\mu_O}{1000} \quad [kJ/mol_{O_2}] \tag{1}$$

$$\log_{10}\left(p_{O_2}\right) = \frac{\Delta G_{onset}\times 10^3}{RT\ln(10)} \tag{2}$$

Here R = 8.314 J mol$^{-1}$ K$^{-1}$, $T$ is the absolute temperature in Kelvin, and p° = 1 bar is the standard gas pressure in **Equation (1)**. O denotes elemental oxygen on a per-mole-of-O-atoms basis; $O_2$ denotes molecular oxygen on a per-mole-of-$O_2$ basis. $\Delta\mu_O = \mu_O - \frac{1}{2}\mu^\circ_{O_2}(T)$ is the oxygen chemical potential relative to half the standard-state $O_2$ chemical potential, expressed in J mol$^{-1}$ O. This reference-state subtraction is essential before converting a database chemical potential to an Ellingham oxygen-potential equivalent. $\Delta G_{onset}$ is calculated from the equilibrium oxygen activity/partial pressure and reported in kJ mol$^{-1}$ $O_2$; the factor of 1000 converts joules to kilojoules. Throughout **Equation (2)**, $pO_2$ is understood as $pO_2/p^\circ$ inside the logarithm. The need for a common elemental reference is illustrated by the SGTE pure-element assessment, which expresses Gibbs energies relative to a defined standard element reference. Reference-state consistency must be checked before comparing oxygen potentials from different thermodynamic expressions [37].

The final checkpoint was assembled using several complementary thermodynamically projected extraction routes. The role of each extraction route is documented in **Supporting Table S1**. All routes used global Gibbs free energy minimization, P = 1 bar, and the same set of 21 temperatures spanning 500-1500 °C. The adaptive phase-boundary scan supplied the full ordered sequence of phase changes. An independent onset-only equilibrium search was applied where the phase-boundary scan returned a boundary-index-0 flag indicating that the scan had not returned a physical boundary. It classified a condensed phase as oxide-bearing when its phase amount exceeded $10^{-8}$ and its oxygen site fraction satisfied $X_O \geq 0.1$. A property-diagram scan (an equilibrium calculation sampled along the oxygen-potential axis) with 100 divisions first bracketed the metal-to-oxide transition. That bracket, an interval containing a change from metal-only to oxide-bearing equilibrium, was refined by 20 bisection iterations that successively halved that interval, with a 25-point direct-equilibrium grid used if the property diagram did not provide a bracket.

Extra calculations improved coverage and recovered difficult regions. Binary and ternary composition sections, containing at most two and three metal components, respectively, were sampled at 5 at.% spacing and quaternary sections containing at most four metal components at 10 at.% spacing using the same phase-boundary scan. Co-bearing binary and ternary cases were recomputed with the upper oxygen-potential limit extended from -100 to $-44$ kJ mol$^{-1}$ O to capture high-temperature CoO onset. Twelve non-converged Fe–Mn–Ni compositions were recomputed with 1000 coarse $\mu$(O) steps, instead of 200, and a longer timeout to resolve narrow Mn-oxide stability windows. Finally, failed or truncated phase strings were identified by the valid-label test, their affected composition–temperature records were cleared, and the equilibrium calculation was rerun. Per temperature checkpointing, bounded worker timeouts, and retry/requeue logic protected already converged records but did not alter thermodynamic values.

The $H_2/H_2O$ equilibrium defines the effective oxygen activity in the HyDR reactor atmosphere. For the oxidation of hydrogen,

$$2\mathrm{H_2} + \mathrm{O_2} \rightarrow 2\mathrm{H_2O}, \Delta G^0_{H2O} \approx -483{,}600 + 88T_K \quad (3)$$

where $T_K$ is temperature in Kelvin, and ΔG° ($H_2O$) is in $\mathrm{J/mol_{O_2}}$. At a specified $H_2/H_2O$ molar ratio r, the effective Gibbs energy of the chemical reaction is:

$$\Delta G_{HyDR}(T, r) = \Delta G^0_{H_2O} - 2RT_K ln(r)\left[J/mol_{O_2}\right] \quad (4)$$

Oxide reduction by $H_2$ is thermodynamically favorable when $\Delta G_{onset} > \Delta G_{HyDR}$, i.e., when the oxide stability line lies above the $H_2/H_2O$ equilibrium line on the Ellingham diagram. We define the reduction driving force as:

$$\Delta G_{rf} = \Delta G_{onset} - \Delta G_{HyDR}\left[kJ/mol_{O_2}\right] \quad (5)$$

A positive $\Delta G_{rf}$ indicates thermodynamically favorable reduction; a larger value implies a larger thermodynamic driving force. The FeO reference line [5],

$$\Delta G_{FeO} \approx \frac{-544{,}000 + 134T_K}{1000}\left[kJ/mol_{O_2}\right] \tag{6}$$

serves as the classical benchmark for iron ore reduction thermodynamics.

The standard Gibbs free energy values for water formation from **Equation (3)** and **(4)** are expressed in joules per mole of $O_2$, whereas onset energies and driving-force plots use kilojoules per mole of $O_2$. The gas result must therefore be divided by 1000 before subtraction in **Equation (5)**, and the onset value multiplied by 1000 before use with R = 8.314 J $mol^{-1}$ $K^{-1}$ in **Equation (7)**. Gas ratios are dimensionless partial-pressure ratios under the ideal-gas approximation. The linear water and FeO expressions are screening approximations; their approximation error relative to temperature-dependent standard-state thermochemical functions is not quantified here, which limits the precision of the calculated humidity thresholds. The iron-oxide reduction literature likewise treats gas composition and temperature as coupled thermodynamic variables, but distinguishes equilibrium condition from rate and reduction degree [5].

The exponential conversion makes gas thresholds sensitive to thermodynamic error. At 1100 °C, an illustrative onset-energy uncertainty of ±10 kJ $mol^{-1}$ $O_2$ gives a multiplicative threshold factor of about 1.55 through $\exp\left(\frac{10^3}{2RT}\right)$. This is an analytical sensitivity example, not a measured database uncertainty or confidence interval. Neither numerical boundary refinement nor surrogate prediction error establishes the accuracy of the underlying assessed Gibbs energies. Grid sampling, phase selection, incomplete coverage and database extrapolation remain separate sources of uncertainty. Bayesian assessment methods implemented in the Extensible Self-optimizing Phase Equilibria Infrastructure (ESPEI) provide a route to propagating thermochemical and phase-equilibrium constraints into parameter uncertainty. Applying that approach here would require the underlying assessment data and parameter distributions; the illustrative gas-ratio sensitivity is not a substitute for such an assessment [38].

### 2.3 Statistical models and predictive assessment

In the statistical analysis, higher-order compositions have four or five nonzero metal fractions. A surrogate model is a fitted statistical approximation to the calculated onset energy. Composition-grouped validation keeps all temperatures for a given alloy together when separating training data from held-out data; a fold is one such held-out subset. Root-mean-square error (RMSE) and mean absolute error (MAE) measure prediction errors in kJ $mol^{-1}$ $O_2$, while the coefficient of determination ($R^2$) measures the reduction in squared prediction error relative to predicting the response mean. These are model-performance measures, not experimental uncertainty estimates. The separation of predictor fitting from predictive assessment follows classical cross-validation. Keeping all temperatures of one composition together applies the broader principle of blocking dependent observations, so a model is not credited for predicting another temperature of an alloy already present in its training fold [39,40].

The higher-order models use 19,817 valid onset records with at least four nonzero metal fractions: 15,351 quaternary and 4,466 quinary records. The response is the stored onset oxygen-potential equivalent in kJ $mol^{-1}$ $O_2$. Metal fractions sum to unity and the reduced temperature is $\tau$ (in °C ) = (T − 1000)/500. The linear mixture–process model contains the five metal fractions and their products with $\tau$ and $\tau^2$; the pair model adds the ten distinct products of two metal fractions. Features are centered and scaled using training-set statistics, and a ridge penalty, proportional to the sum of squared regression coefficients, stabilizes the compositionally constrained design matrix. Standardized pair coefficients describe this fitted basis and are not CALPHAD interaction parameters or independently measured pair energies. The penalty follows the ridge-regression treatment of correlated predictors; its use is a numerical regularization choice rather than evidence for a particular atomistic interaction [41]. The use of metal fractions and pair products also follows the logic of mixture-response models, in which the predictors are proportions constrained to sum to one. Compositional-data theory explains why their covariance and coefficient interpretation differ from those of independently variable concentrations. Here, the polynomial basis accommodates the binary and pure-element limits directly; logarithmic composition ratios would require separate treatment of zero fractions [42,43].

We used five folds by composition, keeping every temperature of an alloy in the same fold. The ridge penalty was selected from the same composition-grouped folds (details in the Supporting Information). Because selection and error use the same folds, scores are selection-conditioned, not independent. Reported $R^2$, RMSE and MAE are calculated from the assembled held-out predictions. Local +5 at.% contrasts preserve composition closure by compensating the other constituents; they are finite perturbations, not unconstrained partial derivatives. Varma and Simon  showed why using the same cross-validation results for hyperparameter selection and error estimation can produce optimistic scores, motivating the explicit distinction between selection-conditioned and nested estimates here [44].

The nonlinear comparison uses histogram gradient boosting, an ensemble of decision trees fitted successively to improve prediction errors using binned input values, with the five fractions and τ as inputs. Five outer composition-grouped folds estimate prediction error; three inner grouped folds select among four combinations of maximum leaf count, minimum samples per leaf and $L^2$ (squared-coefficient) regularization penalty: (15,30,1), (31,30,1), (15,60,3) and (31,60,5). Fixed settings are squared-error loss, learning rate 0.06, 220 iterations, 128 bins, no early stopping (training runs for the specified number of iterations) and random seed 20260731. These tests assess interpolation of calculated onset values at unseen compositions within the sampled domain. Not that they neither validate TCOX10 against experiment nor establish extrapolation to other alloy systems or temperatures. Gradient boosting follows the sequential fitting of additive predictive functions described by Friedman [45]. Its predictive flexibility is useful for a curved thermodynamic response surface, but does not turn a fitted feature effect into a causal or thermodynamic interaction parameter.

## 3 Results and discussion

The results establish the HyDR operating window as a function of oxide composition, temperature, and phase-boundary complexity. The analysis first examines how the stability of oxide phase changes with alloying elements and then compares the predicted thermodynamic results with published experimental observations to distinguish thermodynamic limitations from the transport and kinetic effects that influence their realization.

### 3.1 Hydrogen reduction operating window and composition dependence

The $H_2/H_2O$ ratio (r) sets the reduction potential and hence can be used to predict the hydrogen-based direct reduction (HyDR) window (**Figure 1** and **Table 1**). **Figure 1(a)** compares the reduction potential equivalents at r = 1, 3 and 10 with the FeO reference. We adopt $H_2/H_2O$ = 3 (75 vol.% $H_2$, 25 vol.% $H_2O$) as the reference atmosphere throughout. This ratio lies just above the FeO/Fe equilibrium at the lowest temperature considered ($r_{min}$, FeO ≈ 2.6 at 700 °C; margin of 2.1 kJ $mol^{-1}$ $O_2$). It is therefore the minimum gas quality that reduces wüstite across the whole 700–1300 °C range. It also represents the moderately humid conditions expected at the reacting interface once product water accumulates. Compositions with $r_{min} < 3$ are thus easier to reduce than iron oxide under the same atmosphere, and those with $r_{min} > 3$ need drier gas. The results can be rescaled to any other ratio with **Equation (4)** and (**7**). At 700 °C and $r = 3$, ΔG° for the gas line and FeO is −415.7 and −413.6 kJ $mol^{-1}$ $O_2$, respectively, giving a small positive FeO reduction margin of 2.1 kJ $mol^{-1}$ $O_2$. This margin increases to 15.0 at 900

°C, 27.8 at 1100 °C and 34.3 kJ mol$^{-1}$ $O_2$ at 1200 °C. Heating therefore widens the FeO equilibrium window under this fixed gas ratio; it does not establish that every multicomponent oxide is reducible.

The minimum $H_2/H_2O$ ratio, $r_{min}$, required for complete reduction of an oxide mixture is the gas ratio at which the oxygen potential of the $H_2/H_2O$ atmosphere equals the onset oxygen potential of the most stable residual oxide. In other words, the reduction driving force is zero ($\Delta G_{rf} = 0$ in **Equation (5)**). Substituting **Equation (4)** into **Equation (5)** and solving for $r$ gives:

$$r_{min} = \exp \frac{\Delta G^0_{H2O}(T) - 1000\Delta G_{onset}}{2RT_K} \tag{7}$$

For $r > r_{min}$, $\Delta G_{rf} > 0$ and the metallic state is the final equilibrium state, whereas for $r < r_{min}$ at least one oxide remains stable. $r_{min}$ is an equilibrium quantity at the reacting oxide/gas interface and contains no kinetic information. The median composition-specific equilibrium threshold, $r_{min}$, differs strongly among the ternary systems, as shown in **Figure 1(b).** At 1100 °C, the $r_{min}$ values are approximately 0.155 for Co–Fe–Ni, 413 for Fe–Mn–Ni, 356 for Cr–Fe–Ni, 1,405 for Co–Fe–Mn and 2,529 for Cr–Fe–Mn. A process ratio above the median threshold does not guarantee reduction of every oxide composition: each feedstock must be compared with its own $r_{min}$. Increasing $r$ from 3 to 10 lowers the gas oxygen-potential equivalent by 19.5 kJ mol$^{-1}$ $O_2$ at 700 °C and 29.5 kJ mol$^{-1}$ $O_2$ at 1200 °C. The increased reduction potential ($r$ = 3 to 10) can make near-threshold compositions reducible but is insufficient for the much larger deficits in Mn- and Cr-bearing regions.

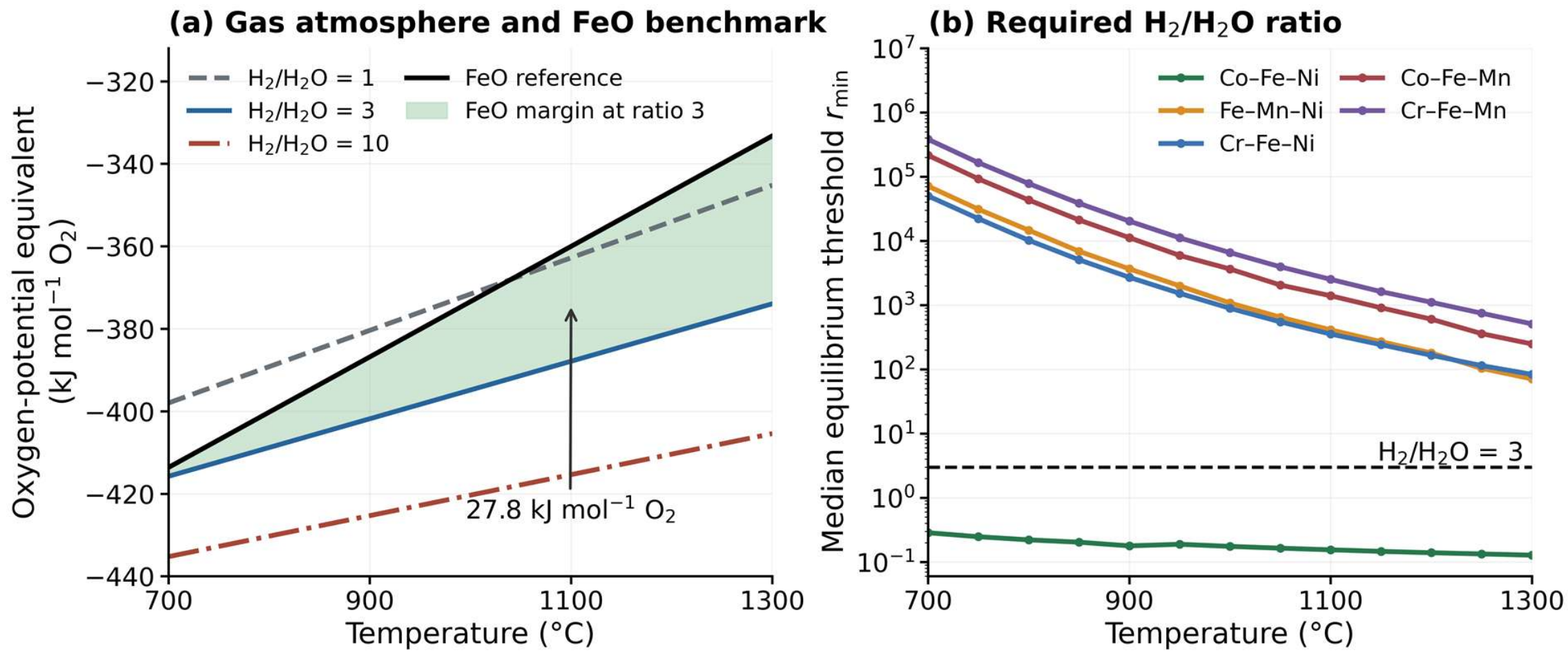

**Figure 1**: **Hydrogen-based direct reduction (HyDR) operating window for mixed oxide reduction**. (a) $H_2/H_2O$ oxygen-potential equivalents at ratios ($r$) 1, 3 and 10 and the FeO reference. Green shaded region marks the reduction window for FeO at $r$ = 3. (b) Median composition-specific equilibrium $H_2/H_2O$ thresholds for the five ternary domains, calculated from the stored boundary-1 energies at each temperature; their binary and pure-element limits are included. These are medians of sampled onset records, not thresholds guaranteeing reduction of an entire domain. At a given composition, $r = r_{min}$ gives zero driving force and $r > r_{min}$ gives a positive margin. The horizontal dashed line is $r = 3$. Temperature-dependent valid coverage and the use of grid samples must be considered when interpreting the curves.

The operating-window interpretation is consistent with established FeO and NiO reduction studies. Spreitzer and Schenk distinguish thermodynamic gas requirements from kinetic limitations in hydrogen reduction of iron oxides [5]. Richardson *et al.* directly followed NiO reduction by X-ray diffraction (XRD) and examined the influence of water vapor on the reduction behavior [46]. These comparisons support the use of local $H_2/H_2O$ as a control variable, but neither supplies a quantitative validation of the much larger thresholds predicted for Mn- or Cr-bearing mixed oxides.

**Table 1**: HyDR thermodynamic operating window: $\Delta G_{HyDR}(kJ/mol_{O_2})$ and $log_{10}(p_{O_2})$ at three $H_2/H_2O$ ratios and key temperatures. The FeO reference from **Equation (6)** is shown for comparison.

| | $\Delta G_{HyDR}(kJ/molO_2)$ | | | FeO ref. (kJ/mol) | $log_{10}(p_{O_2})$ | | |
|---|---|---|---|---|---|---|---|
| $T(^0C)$ | $r = 1$ | $r = 3$ | $r = 10$ | | $r = 1$ | $r = 3$ | $r = 10$ |
| 700 | -398.0 | -415.7 | -435.2 | -413.6 | -21.36 | -22.31 | -23.36 |
| 800 | -389.2 | -408.8 | -430.3 | -400.2 | -18.94 | -19.90 | -20.94 |
| 900 | -380.4 | -401.8 | -425.3 | -386.8 | -16.94 | -17.89 | -18.94 |
| 1000 | -371.6 | -394.8 | -420.3 | -373.4 | -15.24 | -16.20 | -17.24 |
| 1100 | -362.8 | -387.8 | -415.3 | -360.0 | -13.80 | -14.75 | -15.80 |

| | $\Delta G_{HyDR}(kJ/molO_2)$ | | | FeO ref. (kJ/mol) | $log_{10}(p_{O_2})$ | | |
|---|---|---|---|---|---|---|---|
| $T(^0C)$ | $r = 1$ | $r = 3$ | $r = 10$ | | $r = 1$ | $r = 3$ | $r = 10$ |
| 1200 | -354.0 | -380.9 | -410.4 | -346.6 | -12.55 | -13.50 | -14.55 |

The interpolated onset maps for Fe–Mn–Ni, Cr–Fe–Mn, Cr–Fe–Ni, Co–Fe–Ni and Co–Fe–Mn are reported in **Supporting Figures S1–S5**. They show the same separation captured by the combined signed driving-force map below: Co–Fe–Ni is readily reduced at $H_2/H_2O$ = 3, while Mn-rich and Cr-rich regions require substantially drier gas (higher reduction potential).

The reduction driving force $\Delta G_{rf}$ (**Equation 5**) varies across the five Fe-containing ternary composition domains at 700, 900, 1100 and 1300 °C under $H_2/H_2O$ = 3 (**Figure 2**). A ternary domain is the triangular section of the five-element composition space obtained by setting the other two element contents to zero; its edges and vertices are the corresponding binary and pure-element limits, respectively. Each point specifies an overall metal composition, rather than a physical surface of a specimen. At each displayed temperature, Co–Fe–Ni has positive driving force throughout the sampled domain. The other four domains contain broad regions with negative driving force, especially toward Mn-rich or Cr-rich compositions. The fraction of sampled points above zero is a grid-based composition statistic, not a reduction degree or a phase fraction. A positive local margin therefore establishes equilibrium feasibility only for that composition and temperature; it cannot be generalized to the entire ternary domain. To compare the domains as a whole, we use the median $\Delta G_{rf}$ of all sampled compositions in each ternary domain at each temperature (5 at.% grid). The median is the value that half of the sampled compositions exceed. It therefore describes the typical composition of a domain, unlike a single favorable point.

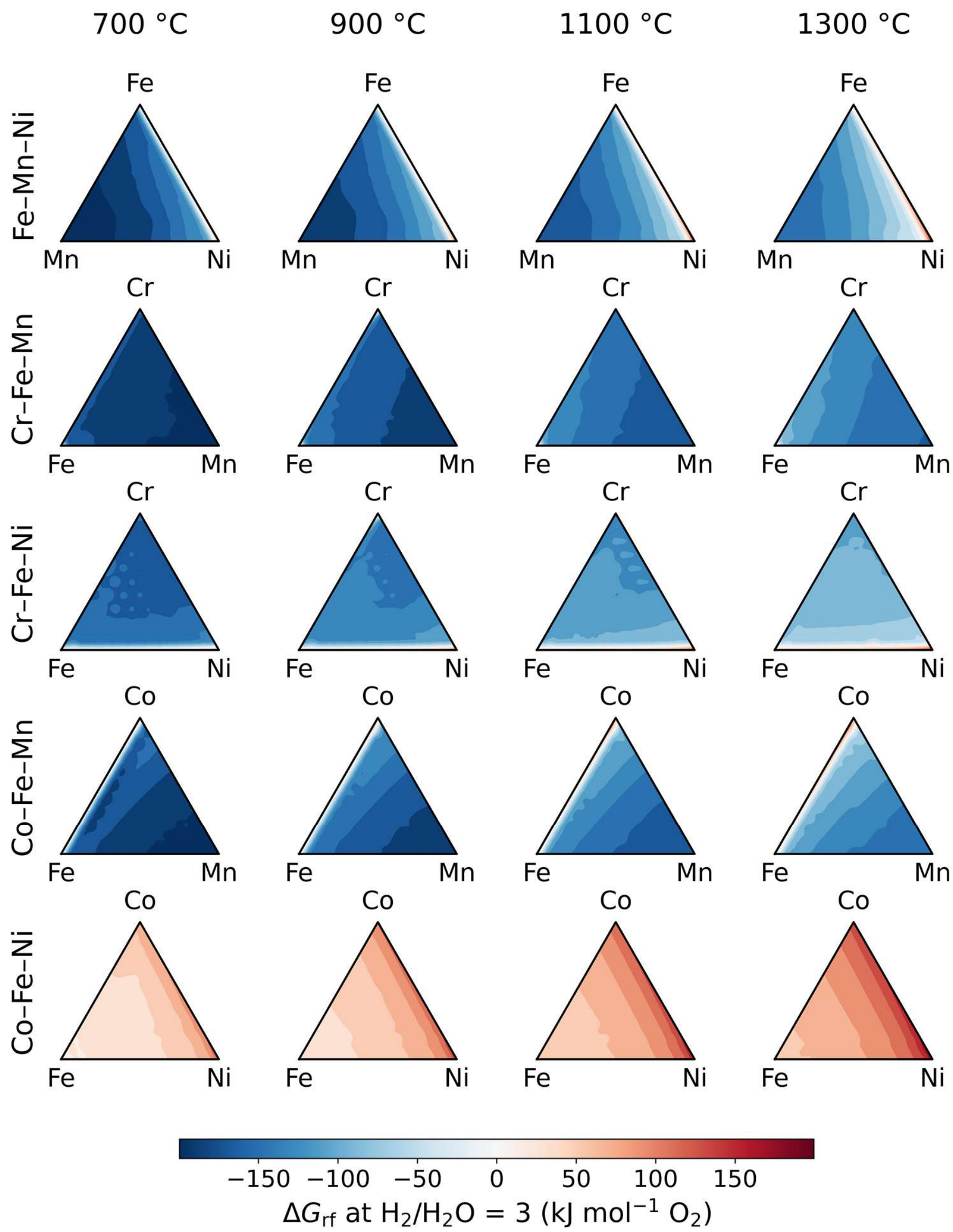

**Figure 2**: **Thermodynamic driving force for ternary oxide mixtures reduction**. Reduction driving force $\Delta G_{rf}$ ($\mathrm{kJ/mol_{O_2}}$) at $H_2/H_2O = 3$ across all five Fe-bearing ternary sub-systems and four temperatures. Ternary systems are arranged by row (Fe–Mn–Ni, Cr–Fe–Mn, Cr–Fe–Ni, Co–Fe–Mn and Co–Fe–Ni) and temperatures by column. A single symmetric colour scale is used for every panel: red denotes $\Delta G_{rf} > 0$ (thermodynamically reducible), blue denotes $\Delta G_{rf} < 0$, and white is the zero-driving force threshold. Co–Fe–Ni is positive throughout the sampled face, whereas the other four sub-systems contain large negative-driving-force regions at $H_2/H_2O = 3$.

From 700 °C to 1300 °C, the median $\Delta G_{rf}$ changes from -163 to -82 kJ/mol $O_2$ for Fe–Mn–Ni, from -190 to -134 kJ/mol $O_2$ for Cr–Fe–Mn, from -157 to -87 kJ/mol $O_2$ for Cr–Fe–Ni, and from -181 to -115 kJ/mol $O_2$ for Co–Fe–Mn. Cr–Fe–Mn is therefore the least favorable face by its median at every displayed temperature, while the positive fraction remains small because only a narrow region near the less reactive endmembers crosses the gas line. Temperature shifts distributions toward positive, but Fe-Mn-Ni and Cr-Fe-Ni swap median order. The emerging practical consequence is that temperature and composition are not interchangeable control variables: heating a poorly selected oxide blend improves its margin, whereas moving away from Mn- or Cr-rich ends can change both the magnitude and the identity of the bottleneck.

The Cr-bearing maps can be compared with the HyDR of Cr–Fe–Ni oxides for stainless-steel synthesis reported by Al-Buainain and Dunand [47]. Their observation of Fe-Ni alloy formation and Cr incorporation into the Fe–Ni matrix is consistent with the role of metallic component activity in determining the reduction boundary. However, their alloy compositions in wt.% (Fe–18Cr–8Ni and Fe–18Cr–35Ni), and their experiments follow a finite-temperature processing path, whereas the maps in this study use metal atomic fractions and represent equilibrium onset states. The agreement is therefore mechanistic rather than a pointwise; a quantitative comparison requires conversion of the reported compositions and knowledge of the interfacial $H_2/H_2O$ ratio during the experiments [19,47].

For the Fe–Mn–Ni composition domain, this ratio gives the required local gas ratio. Its median decreases from $7.16 \times 10^4$ at 700 °C to 71.2 at 1300 °C, while the maximum remains 1326 at the higher temperature. The corresponding Co–Fe–Ni median is below unity at every displayed temperature. Heating lowers the gas-ratio requirement but does not remove the Mn-controlled reduction bottleneck. The exponential conversion in **Equation (7)** amplifies apparently moderate differences in onset energy. At 1100 °C, the median $r_{min}$ values are 413 for Fe–Mn–Ni, 2529 for Cr–Fe–Mn, 356 for Cr–Fe–Ni, and 1405 for Co–Fe–Mn, compared with 0.16 for Co–Fe–Ni. In a feed containing 5 vol.% $H_2$, these median thresholds correspond to approximately 121, 20, 140, and 36 ppm $H_2O$ by mole, respectively, under the ideal-gas convention used here. The values are equilibrium targets rather than permissible inlet specifications: gas consumption, oxygen removal from the specimen, leaks, and axial gradients can raise the local water activity above the measured inlet value. Nevertheless, the calculation explains why conditions all described informally as “dry hydrogen” can produce varying reduction degrees in different laboratories.

Local water accumulation provides a physical reason to distinguish the interfacial threshold from an inlet specification. El-Zoka *et al.* detected heavy-water formation in nanopores and deuterium partitioning during reduction of iron oxide with deuterium at 700 °C [48]. The isotope-tracing experiment does not measure the present mixed-oxide thresholds, but it shows that the water can be trapped inside pores during an evolving porous microstructure and can shift the local reduction potential.

## 3.2 Oxide phase selection and temperature dependence

The dominant oxide phase at onset boundaries in Fe-bearing compositions is HALITE (Mn, Ni)O rock-salt (42.4% of valid boundary events), followed by CORUNDUM $Cr_2O_3$-type (28.6%) and SPINEL ((Co, Fe)(Co, Cr, Fe, Mn)$_2$O$_4$-type, 27.6%)). A minor cubic A13-type oxide phase (CUB_A13, 1.5%) appears in Mn-bearing compositions between 500 °C and 1150 °C. The elevated CORUNDUM fraction compared to earlier partial-dataset analyses reflects the successful gap-fill of Cr-bearing compositions (in particular Cr–Fe binary and Cr–Fe–Mn, Cr–Fe–Ni faces) where $Cr_2O_3$ onset is systematically captured by the expanded dataset.

The phase-resolved onset distributions also show why a structural label cannot determine the reducing atmosphere. Grouping the valid Fe-bearing records by the onset phases identified in **Figure 3,** the median onset energies at 700 °C are -593.4 kJ/mol $O_2$ for SPINEL, -588.3 for HALITE and -573.1 for CORUNDUM. Relative to the $r = 3$ gas line, the corresponding median driving forces are -177.7, -172.6 and -157.3 kJ mol$^{-1}$ $O_2$. At 1100 °C, the median onset energies increase to -525.0, -514.1 and -498.2 kJ/mol $O_2$, respectively, but the corresponding driving forces remain negative at -137.2, -126.3 and -110.4 kJ mol$^{-1}$ $O_2$. Heating improves these pooled medians without making the median composition in any of the three phase groups reducible at $r = 3$. The positive driving-force tails and the readily reducible Co–Fe–Ni compositions must therefore be distinguished from the median behavior of the pooled Fe-bearing population.

The smoothed first-oxide fields across the five Fe-bearing ternary faces identify the local controlling structures (**Supporting Figures S6–S10**). HALITE controls the Mn-bearing faces, CORUNDUM controls most of Cr–Fe–Ni, and SPINEL occupies inter mediate Co-, Fe-, Mn, and Cr-bearing regions. These categorical maps remain in Supporting Information because the manuscript uses the phase-frequency summary in **Figure 3** to carry the reduction argument.

Temperature changes the frequency with which each oxide structure controls the oxidation-onset boundary across the four Fe-bearing ternary systems summarized in **Figure 3**: (a) Fe–Mn–Ni, (b) Co–Fe–Mn, (c) Cr–Fe–Mn and (d) Cr–Fe–Ni. Each stacked bar is normalized to the valid onset events for that system and temperature. The percentages therefore describe the distribution of phase identities across the sampled compositions; they are not equilibrium mass or volume fractions in a single specimen. Under the onset classification used here, a coexisting assemblage is assigned to its first-listed phase. In the reverse, equilibrium reduction direction, this first oxide encountered from the metallic side identifies the final oxide-bearing boundary that must be crossed to reach the modeled metallic state. Thus, **Figure 3** indicates how widely a particular class of reduction bottleneck occurs, rather than how much residual oxide a specimen will contain.

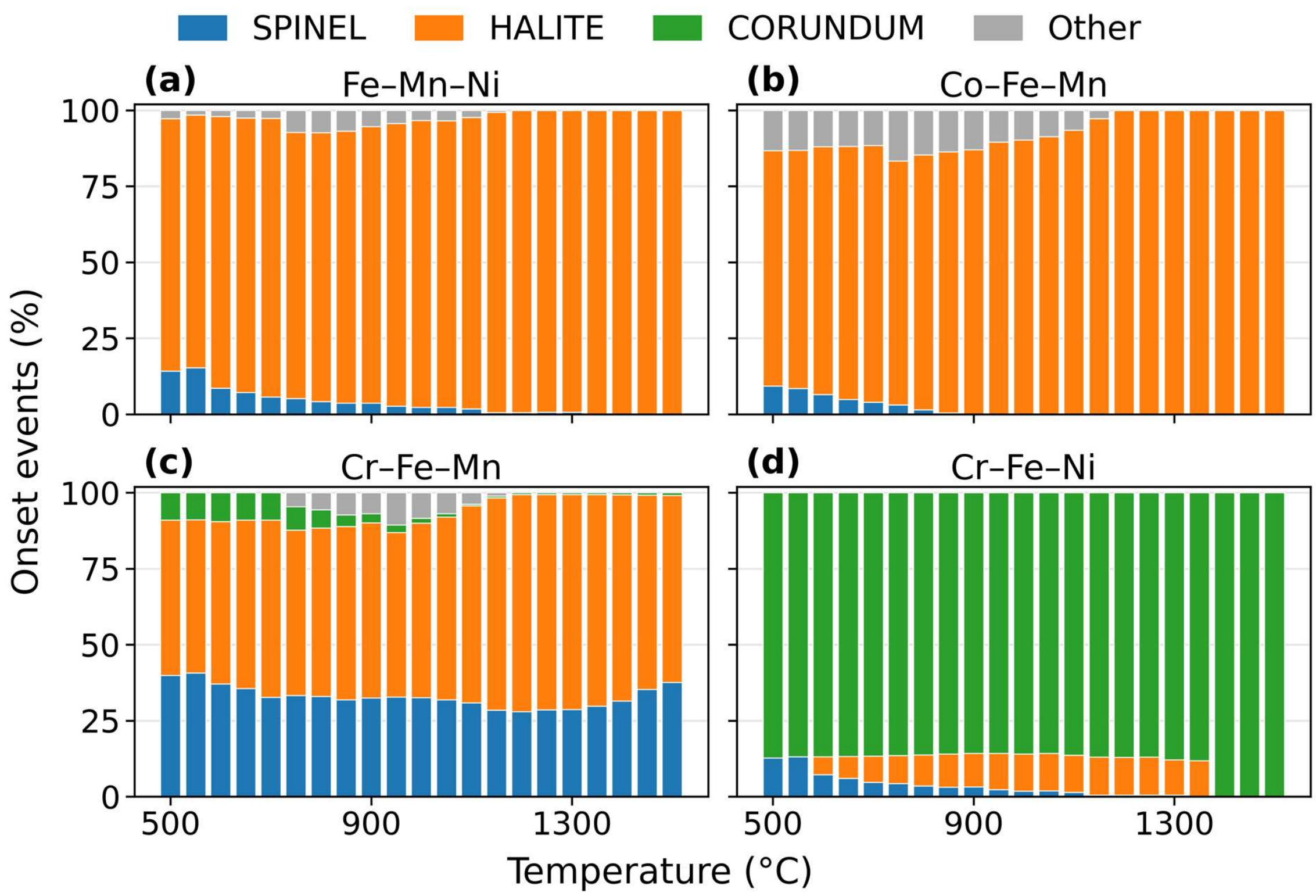

**Figure 3**: **Temperature-dependent distribution of oxidation-onset phases across ternary alloy systems**. Phase distribution of oxidation-onset events as a function of temperature in the (a) Fe–Mn–Ni, (b) Co–Fe–Mn, (c) Cr–Fe–Mn and (d) Cr–Fe–Ni systems. Stacked bars showing the percentage of valid oxidation-onset events assigned to each oxide phase at each temperature. Each bar is normalized within its ternary system and temperature; percentages represent phase-identity frequencies across sampled compositions, not mass or volume fractions within a specimen. HALITE becomes nearly exclusive at high temperature in (a) Fe–Mn–Ni and (b) Co–Fe–Mn. (c) Cr–Fe–Mn retains a substantial SPINEL contribution, whereas (d) Cr–Fe–Ni is predominantly CORUNDUM-controlled. The grey 'Other' category groups all remaining onset phases. Coexisting onset assemblages follow the first-listed-phase convention described in the methodology. The phase distribution identifies the most frequent final equilibrium reduction bottlenecks; their $H_2/H_2O$ requirements must be obtained from the corresponding onset energies.

Fe–Mn–Ni and Co–Fe–Mn show the clearest shift toward HALITE control (**Figure 3(a)** and **(b)**). In Fe–Mn–Ni, HALITE accounts for roughly four-fifths of onset events at 500 °C and approaches the entire classified population at the upper end of the temperature range. The SPINEL contribution falls from about 10–15% at the lowest temperatures to only a few percent around 900–1100 °C, while the other category (grey) disappears entirely at higher temperatures (>1200 °C). Co–Fe–Mn exhibits a comparable progression: HALITE increases from approximately three-quarters of the events at 500 °C to essentially all events from about 1200 °C onward, with SPINEL becoming negligible by approximately 900 °C. These approximate proportions are read from the plotted bars. The common trend is consistent with the persistence of Mn-bearing monoxide solid solutions near the metal/oxide boundary [20,21,49] although the HALITE label alone does not establish the cation composition at an individual point. The loss of SPINEL from the onset statistics does not imply that SPINEL cannot form earlier during reduction at higher oxygen potential.

HyDR experiments provide a useful comparison with this Mn-bearing HALITE bottleneck. Shankar *et al.* studied reduction behavior of multicomponent oxide precursors targeting equiatomic Co–Fe–Mn–Ni alloy and found that mechanically mixed and pre-sintered precursors both reached approximately 80% reduction after isothermally holding for 1 h at 700 °C, leaving MnO alongside metallic Co, Fe and Ni; a small amount of Fe was retained in the (Fe,Mn)O solid solution [20]. Their subsequent in situ high-energy X-ray diffraction study resolved precursor-dependent pathways: mechanically mixed powders reduced through HALITE, SPINEL and $Mn_3O_4$ transient intermediates, whereas the pre-sintered multicomponent oxide reduced into metallic phases and MnO through a different sequence [21]. These observations support the distinction between transient oxides and a persistent Mn-rich monoxide residue. Because the experiments concern a selected quaternary composition and finite-time reduction conditions, they support the proposed phase chemistry but do not quantitatively validate the ternary onset percentages or their temperature dependence.

Cr–Fe–Mn differs from the other Mn-bearing faces because SPINEL remains a substantial onset phase throughout the plotted interval (**Figure 3(c)**). Its share is approximately 30–40% at lower temperatures, decreases to roughly one-quarter near 1200 °C, and rises again toward 1500 °C. HALITE remains the largest category, while CORUNDUM contributes mainly at low temperature and becomes small above approximately 1000–1100 °C. This persistent SPINEL population is consistent with competition between Cr-bearing SPINEL solutions and Mn-bearing monoxides across different compositions. The simultaneous appearance of several colors in a bar represents different controlling phases at different sampled compositions, rather than coexistence of all those phases in every sample. Consequently, the declining CORUNDUM share cannot be interpreted as removal of the Cr-related reduction constraint: Cr may remain accommodated in other stable oxide solutions, and the local boundary energy is still needed to determine reducibility.

Cr–Fe–Ni, in contrast, remains overwhelmingly CORUNDUM-controlled, as shown by **Figure 3(d)**. CORUNDUM represents approximately 85–90% of the classified onset events over most of the range and approaches the full population at the highest plotted temperatures. The low-temperature SPINEL contribution progressively disappears, while HALITE occupies only a relatively small share. Compared with Cr–Fe–Mn, the replacement of Mn by Ni therefore changes the dominant structural competition, consistent with a broad Cr-rich sesquioxide stability field when the Mn-bearing monoxide pathway is absent. Oxidation experiments by Laplanche *et al*. identified a thin inner $Cr_2O_3$ layer beneath a predominantly $Mn_2O_3$ scale on equiatomic CrMnFeCoNi at 600 °C in air [35]. This observation supports selective Cr-rich oxide formation, but the alloy composition, oxygen potential and oxidation geometry differ from the present ternary reduction boundary. It is therefore qualitative evidence for the phase assignment, rather than a measurement of the calculated CORUNDUM frequency or hydrogen requirement.

Overall, the panels show that heating simplifies the onset-phase distribution most strongly in Fe–Mn–Ni and Co–Fe–Mn, whereas Cr–Fe–Mn retains competing HALITE and SPINEL fields and Cr–Fe–Ni remains predominantly CORUNDUM-controlled. The other category (grey) is appreciable at low and intermediate temperatures in the Mn-bearing systems, but its aggregation prevents individual minor phases from being distinguished from this plot alone. Changes in normalized percentages, particularly near the highest temperatures, must also be interpreted alongside the number and composition coverage of valid calculations. Most importantly, a larger HALITE or CORUNDUM share is not itself a larger reduction driving force. The phase frequencies identify the predicted residual oxide chemistry across the sampled domain; the composition-specific onset energy and $H_2/H_2O$ threshold determine the equilibrium condition for removing that residue.

Oxidation experiments provide a complementary test of phase selection. Yu *et al*. showed that thermodynamics, kinetics and growth mode compete during early oxidation of CoCrFeNi [27], while Gwalani *et al*. resolved chemically differentiated oxide growth in high-entropy alloys [28] . Together with the Mn- and Cr-rich oxide scales reported by Laplanche *et al*. , these results support selective partitioning among oxide structures [35]. They also explain why an equilibrium frequency histogram cannot predict scale-layer thickness, spatial order or growth rate, and why air-oxidation data should not be treated as direct measurements of a hydrogen-reduction endpoint.

### 3.3 Composition sensitivity and surrogate performance

A binary host refers to the two-metal base composition into which a third metal is introduced (**Figure 4**). Pooled regression combines data across the specified compositions and temperatures, whereas a local fit considers only a restricted composition interval. Standardization expresses each variable relative to its mean in units of its standard deviation; accordingly, the standardized coefficient $\beta_k$, defined in **Equations (8)** and **(9)**, quantifies the change in the response on this normalized scale. Because the metal fractions are constrained to sum to unity (composition closure), increasing the fraction of one element necessarily requires a corresponding decrease in one or more other constituents. Therefore, each coefficient represents the effect of adding the third element relative to the associated compositional compensation and reference composition.

All composition sensitivities in this section are expressed as standardized ordinary least-squares (OLS) coefficients, $\beta_k$, where the subscript k identifies the metal whose fraction is varied (e.g., $\beta_{Mn}$ for Mn). The same symbol is used in the text, in **Figures 4** and **5**, and in **Supporting Table S3** and **Supporting Figure S12**. For each data subset, every variable v is first converted to a z-score, $z(v) = (v - \bar{v})/s_v$, where $\bar{v}$ and $s_v$ are the mean and sample standard deviation of v within that subset. The coefficients are then estimated by minimizing the sum of squared residuals, ε, of a linear model. For the addition of a third element k to a binary host A–B (**Figures 4** and **5**), the model is fitted to all valid onset records on the A–B–k ternary face, i.e., over the full range of host ratios and third-element fractions and over all 21 temperatures:

$$z(\Delta G_{\text{onset}}) = \beta_0 + \beta_k z(x_k) + \beta_{\text{T}} z(T) + \varepsilon \tag{8}$$

where $x_k$ is the atomic fraction of k on the metal-only basis. The term $\beta_T$ z(T) removes the common temperature dependence, so $\beta_k$ is a temperature-controlled average over 500–1500 °C rather than a value at one temperature or one host composition. For the pooled and temperature-resolved analyses of the Fe-bearing population, the four metal fractions enter simultaneously, with Ni as the omitted reference because the fractions sum to unity:

$$z(\Delta G_{\text{onset}}) = \beta_0 + \sum_k \beta_k z(x_k) + \beta_{\text{T}} z(T) + \varepsilon, \qquad k \in \{\text{Co}, \text{Cr}, \text{Fe}, \text{Mn}\} \tag{9}$$

The temperature term is omitted when **Equation (9)** is fitted separately at each temperature (**Supporting Figure S12**). $\beta_k$ is dimensionless: it gives the change in $\Delta G_{onset}$, in units of its standard deviation, for a one-standard-deviation increase in $x_k$ at fixed temperature. A negative $\beta_k$ therefore means that adding k lowers the onset oxygen potential, stabilizes the residual oxide and raises $r_{min}$ (**Equation 7**), whereas a positive $\beta_k$ means that k destabilizes the oxide and facilitates reduction. Fitting the same models to $\log_{10} p_{O2}$ instead of $\Delta G_{onset}$ gives the right-hand panel of **Figure 4**. In the sliding-window analysis (**Figure 5**), **Equation (8)** is refitted only to the records whose fraction of the first host element lies within ±15 at.% of each window centre (10–90 at.%, at least 20 records per window), giving a local $\beta_k$ for that host ratio. Because each coefficient is scaled by the spread of its own subset, $\beta_k$ ranks the sign and relative strength of composition effects within a fit. It can be converted to an absolute sensitivity through $\partial\Delta G_{onset}/\partial x_k \approx \beta_k\ s_{\Delta G}/s_{x,k}$, but coefficients obtained from subsets with different composition ranges should not be compared as absolute energies.

The sensitivity analysis is examined at four complementary levels: pooled element effects, temperature dependence, transfer into each binary host, and local changes with host ratio and added-element concentration. The pooled regression established the overall hierarchy: Mn has the strongest oxide-stabilizing effect, followed by Cr, while Fe has a weaker effect and Co shows the smallest pooled effect relative to the Ni reference. These coefficients provide global summaries rather than universal composition derivatives; the analyses presented in **Figure 4** demonstrate why they cannot be treated as universal composition derivatives. The corresponding coefficients and test statistics are reported in **Supporting Table S3**.

Temperature-resolved fits of **Equation (9)**, without the temperature term, were performed separately at each of the 21 temperatures from 500 to 1500 °C (**Supporting Figure S12**). Up to 1400 °C they preserve the pooled hierarchy at every temperature. Between 500 and 1400 °C, the stabilizing Mn coefficient strengthens from $\beta_{Mn} = -0.56$ to $-0.73$, $\beta_{Cr}$ from $-0.42$ to $-0.66$ and $\beta_{Fe}$ from $-0.13$ to $-0.39$, whereas $\beta_{Co}$ remains close to zero ($|\beta_{Co}| \leq 0.07$) throughout. The fit quality increases from $R^2 = 0.32$ at 500 °C to 0.68 at 1400 °C but falls to 0.27 and 0.17 at 1450 and 1500 °C, where fewer valid Fe-bearing records remain (1031 and 893, compared with 2410 at 500 °C); the coefficients in these two slices are therefore not interpreted. Because the standardization and the sampled population change with temperature, these coefficients compare relative effects within each slice; they are not absolute composition derivatives or evidence of a kinetic transition.

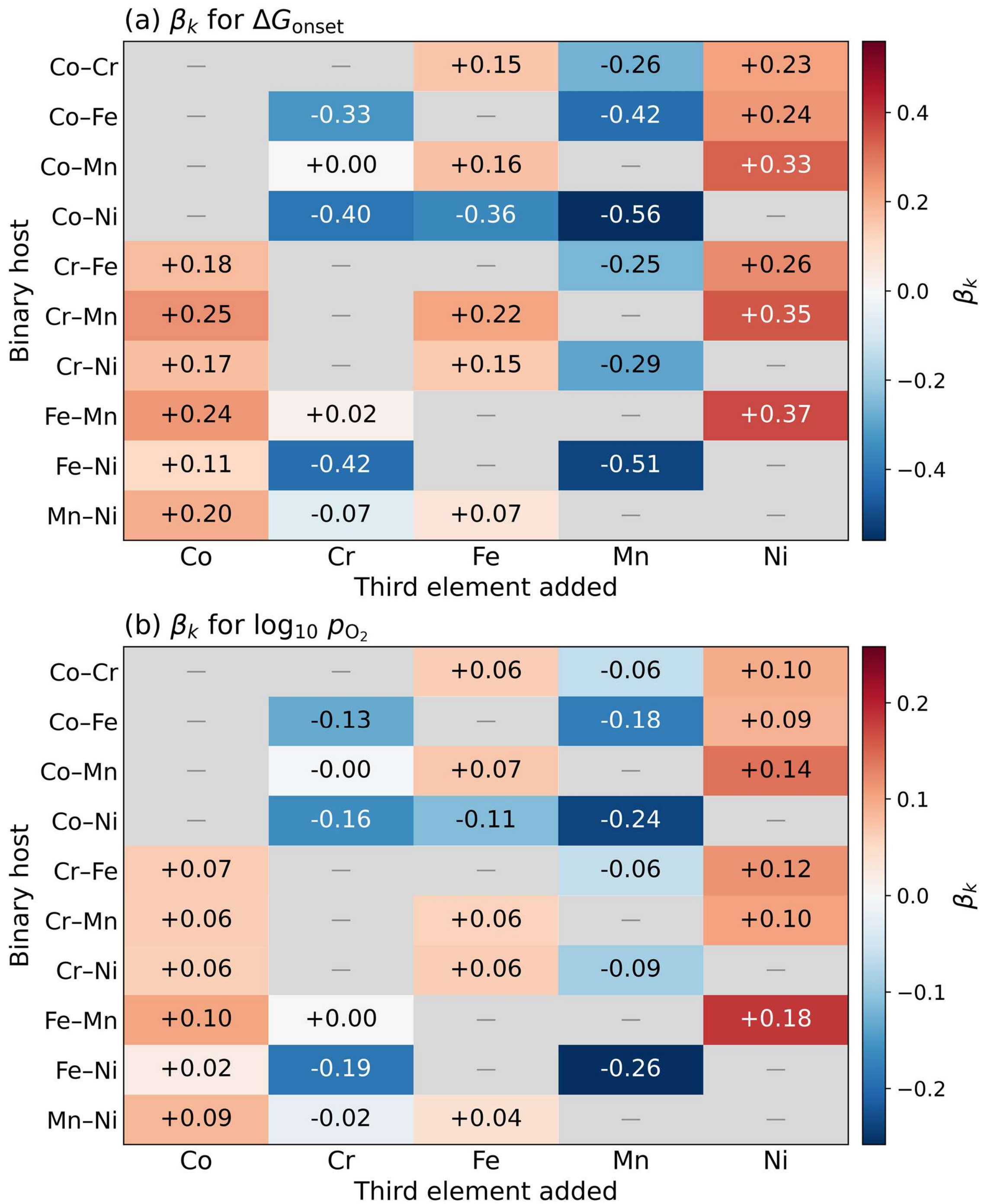

| Binary host | Co | Cr | Fe | Mn | Ni |
|---|---|---|---|---|---|
| Co–Cr | — | — | +0.15 | -0.26 | +0.23 |
| Co–Fe | — | -0.33 | — | -0.42 | +0.24 |
| Co–Mn | — | +0.00 | +0.16 | — | +0.33 |
| Co–Ni | — | -0.40 | -0.36 | -0.56 | — |
| Cr–Fe | +0.18 | — | — | -0.25 | +0.26 |
| Cr–Mn | +0.25 | — | +0.22 | — | +0.35 |
| Cr–Ni | +0.17 | — | +0.15 | -0.29 | — |
| Fe–Mn | +0.24 | +0.02 | — | — | +0.37 |
| Fe–Ni | +0.11 | -0.42 | — | -0.51 | — |
| Mn–Ni | +0.20 | -0.07 | +0.07 | — | — |

| Binary host | Co | Cr | Fe | Mn | Ni |
|---|---|---|---|---|---|
| Co–Cr | — | — | +0.06 | -0.06 | +0.10 |
| Co–Fe | — | -0.13 | — | -0.18 | +0.09 |
| Co–Mn | — | -0.00 | +0.07 | — | +0.14 |
| Co–Ni | — | -0.16 | -0.11 | -0.24 | — |
| Cr–Fe | +0.07 | — | — | -0.06 | +0.12 |
| Cr–Mn | +0.06 | — | +0.06 | — | +0.10 |
| Cr–Ni | +0.06 | — | +0.06 | -0.09 | — |
| Fe–Mn | +0.10 | +0.00 | — | — | +0.18 |
| Fe–Ni | +0.02 | -0.19 | — | -0.26 | — |
| Mn–Ni | +0.09 | -0.02 | +0.04 | — | — |

**Figure 4**: **Third-element effects on oxide stability across binary host systems.** Third-element sensitivity across all ten binary host systems after controlling for temperature. Columns identify the added element and rows identify the binary host. Panel (a) reports standardized coefficients $\beta_k$ (**Equation 8**), fitted over all valid onset records on each ternary face and all temperatures from 500 to 1500 °C, for $\Delta G_{onset}$ and panel (b) reports coefficients $\beta_k$ for $\log_{10}(p_{O_2})$. Negative coefficients (blue) indicate oxide stabilisation and a larger hydrogen-reduction requirement, while positive coefficients (red) indicate oxide destabilization.

The binary-host analysis in **Figure 4** separates the response to a third element from changes in the two host fractions. For a host A–B and an added element k, $\beta_k$ is fitted to all calculations on the A–B–k ternary face (**Equation 8**), i.e., over all A:B ratios, all k contents and all temperatures, rather than at one fixed (e.g., equiatomic) host composition; **Figure 5** then resolves the same coefficient as a function of the A:B ratio. Mn is strongly oxide-stabilizing when added to Co–Ni ($\beta_{Mn}$ = -0.56), Fe-Ni (-0.51), and Co–Fe (-0.42). Cr is likewise stabilizing on Co–Fe, Co–Ni, and Fe–Ni hosts, whereas Ni is oxide-destabilizing on most hosts that do not already contain it. The same sign pattern appears for both $\Delta G_{onset}$ and $\log_{10}(pO_2)$, although the standardized magnitudes differ. The grey cells are chemically inapplicable additions because the nominal third element is already a host constituent.

Local variations along each binary host are resolved with a sliding window, a restricted interval moved successively across the host ratio (**Figure 5**). Several effects vary smoothly in magnitude, but local sign changes occur where the host ratio changes the first stable oxide. For example, Cr added to Co–Mn changes from positive to negative as Co increases, while the Co and Fe effects on Cr–Fe decay towards zero at the Cr-rich end. The strong Mn response on Fe–Ni and Co–Ni remains negative over the full sampled host range. These curves identify composition intervals with changing local sensitivity and show which pooled coefficients are robust versus phase-conditioned.

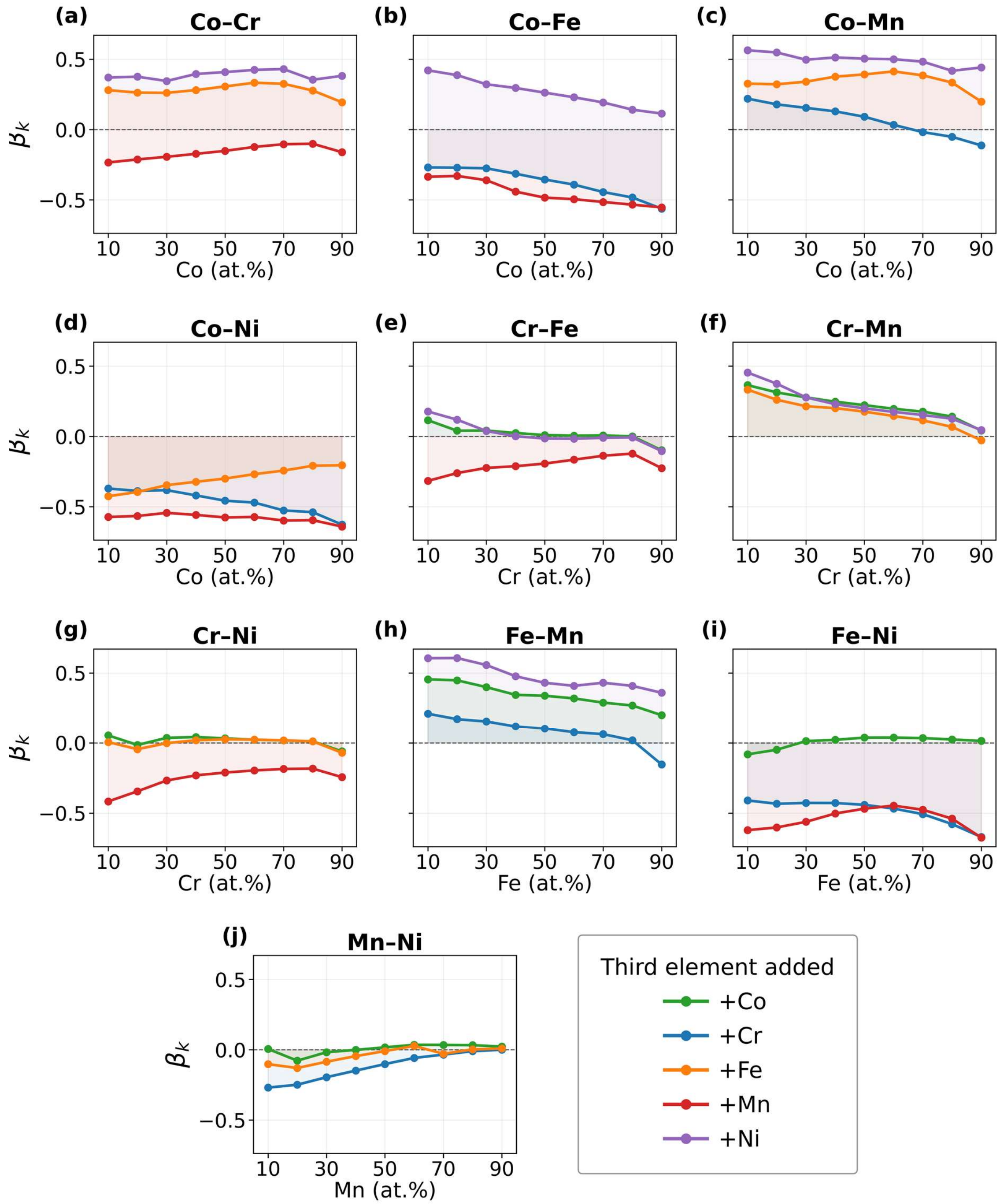

**Figure 5**: **Composition-dependent third-element sensitivity across binary host systems**. Sliding-window third-element sensitivity, $\beta_k$ (Equation 8), as a function of binary host ratio. Panels (a)–(j) show the ten binary hosts; the horizontal axis gives the fraction of the first named host element, and colored curves identify the added element. Coefficients were fitted within ±15 at.% composition windows while controlling for temperature. Crossings of the zero line identify host ratios at which the same addition changes from oxide-stabilizing to oxide-destabilizing, commonly because the controlling onset phase changes.

The concentration-profile analysis distinguishes the finite shift accumulated from a pure-binary baseline from the local derivative at each addition level. Within the practically relevant alloying range (0–30 at.%), which covers typical additions in established alloy classes, e.g., about 3–12 wt.% Mn in medium-Mn steels [25] and about 18 wt.% Cr and 8–10 wt.% Ni in austenitic stainless steels [24], Mn and Cr produce the largest negative changes, and most of their response occurs at the first increments. Co, Fe, and Ni additions are weaker or host-dependent over the same interval. The extreme upturns near pure-element limits reflect sparse end-member phase changes outside the intended alloy-design range; the 0–30 at.% interval therefore provides the relevant domain for interpreting dilute-to-concentrated alloying trends.

The heatmap in **Figure 4** gives the temperature-controlled average response of an added element, whereas **Figure 5** shows how that response changes as the ratio of the two host elements is varied. The two quantities should be read together: a pooled coefficient identifies the net direction over an entire binary edge, while the sliding-window range reveals whether that direction is uniform or results from cancellation across different oxide phase fields. The numerical values below are standardized coefficients and therefore compare effective strength, not absolute reaction energies. Literature comparisons are necessarily qualitative because published oxidation and reduction studies generally examine selected multicomponent alloys rather than the same continuous binary-host composition series.

*Co-Cr host:* Mn is the only consistently oxide-stabilizing addition in the pooled analysis ($\beta_{\mathrm{Mn}} = -0.263$); its local coefficient remains negative from approximately −0.23 to −0.10 across the Co fraction. Fe and Ni are oxide-destabilizing on average (+0.153 and +0.226) and remain positive locally, with ranges of about +0.19 to +0.33 and +0.34 to +0.43, respectively. Thus, changing the Co:Cr ratio alters the magnitude but not the effective stabilization order, Mn < Fe < Ni of $\Delta G_{\mathrm{onset}}$. This agrees qualitatively with oxidation observations in Co-Cr-Fe-Ni alloys, where Cr is sequestered in a CORUNDUM layer while Ni is retained beneath the scale and Fe participates in an outer SPINEL; this elemental partitioning makes Mn addition the more plausible route to a still more stable oxide assemblage [27,50].

*Co-Fe host:* Cr and Mn strongly stabilize the onset oxide, with pooled coefficients $\beta_{\mathrm{Cr}} = -0.327$ and $\beta_{\mathrm{Mn}} = -0.422$. Their local responses span −0.56 to −0.27 and −0.55 to −0.33, respectively, and become most negative toward the Co-rich part of the edge. Ni has the opposite effect ($\beta_{\mathrm{Ni}} = +0.242$; local range +0.11 to +0.42), with the largest benefit toward Fe-rich compositions. The contrast is consistent with the ability of Cr and Mn to enter stable

CORUNDUM, SPINEL, or HALITE oxides, whereas Ni incorporation and subsequent alloy formation can simplify and accelerate Fe-oxide reduction pathways [51]. The calculated coefficients describe equilibrium onset, while the experimentally observed acceleration also contains kinetic contributions from alloy formation and suppression of reoxidation.

*Co-Mn host:* Fe and Ni destabilize the onset throughout the sampled edge: their pooled coefficients are +0.164 and +0.330, and their local ranges are +0.20 to +0.41 and +0.42 to +0.56. Cr is different. Its pooled coefficient is nearly zero (+0.001), but the local curve ranges from +0.22 to -0.11 and crosses zero as the Co fraction increases. The near-zero average therefore represents cancellation between two composition regimes rather than a uniformly negligible Cr effect. This host dependence is compatible with the experimentally established competition between fast-growing Mn-rich oxides and protective Cr-rich CORUNDUM/chromia; changing the cation balance can switch which oxide solution controls onset [35,50,52].

*Co-Ni host:* All three possible additions stabilize the oxide relative to the readily reducible Co–Ni base. Mn is strongest ($\beta_{\mathrm{Mn}} = -0.559$; local range -0.64 to -0.54), followed by Cr (-0.403; -0.63 to -0.37) and Fe (-0.359; -0.42 to -0.20). None changes sign over the binary edge, so Co–Ni is unusually sensitive to even modest contamination by a more reactive oxide former. The result is consistent with the comparatively facile hydrogen reduction of NiO and with the tendency of Ni to remain metallic or alloyed beneath Cr-bearing oxide scales [27,46,53]. Consequently, Co-Ni provides a useful low-stability baseline against which Mn-, Cr-, and Fe-induced increases in the hydrogen requirement can be measured.

*Cr-Fe host:* Mn is consistently stabilizing ($\beta_{\mathrm{Mn}} = -0.254$; local range -0.32 to -0.12). Co and Ni are destabilising in the pooled fit (+0.177 and +0.263), but their local ranges, −0.10 to +0.12 and -0.10 to +0.18, cross zero toward the Cr-rich end. Dilution by Co or Ni therefore helps most on the Fe-rich side and loses effectiveness when Cr-rich CORUNDUM controls onset. Fe-Cr oxide compacts are experimentally known to require coupled reduction and sintering, while alloying Fe with Cr can lower Cr activity and permit reduction at lower temperature than for pure $Cr_2O_3$ [8,19]. The sign-changing curves are consistent with that activity effect becoming insufficient in the Cr-rich limit.

*Cr-Mn host:* Co, Fe, and Ni are all destabilizing in the pooled analysis (+0.255, +0.220, and +0.349). Their local ranges are +0.04 to +0.36, -0.03 to +0.33, and +0.04 to +0.46, respectively. The benefit of each addition decreases toward the Cr-rich side, and Fe approaches a weakly negative response in the most Cr-rich windows. The common positive direction follows from dilution of a binary host composed of two strong oxide formers, but the convergence toward zero shows that concentrated Cr-rich oxide solutions are difficult to displace. Published Cantor-alloy oxidation studies likewise identify Mn-rich external oxides together with Cr-rich inner layers, demonstrating that the two elements can stabilize distinct parts of a multicomponent scale rather than behaving as one additive oxide former [28,35,50].

*Cr-Ni host:* Mn remains stabilizing across the full edge ($\beta_{Mn} = -0.295$; local range -0.42 to -0.18). Co and Fe are mildly destabilising on average (+0.169 and +0.147), yet their local responses lie close to zero (-0.06 to +0.05 and -0.07 to +0.03). Thus, the pooled positive coefficients overstate their reliability as design levers at specific Cr:Ni ratios, whereas the Mn penalty is robust. The result is consistent with Cr sequestration in CORUNDUM and Ni retention beneath the oxide in CoCrFeNi, as well as the formation of chromia plus mixed $(Co,Ni)Cr_2O_4$ SPINEL in CrCoNi [27,50]. Mn addition introduces another strong oxide former and therefore preserves a negative response even where Ni is abundant.

*Fe-Mn host:* Ni and Co destabilize the onset oxide, with pooled coefficients of +0.370 and +0.243 and local ranges of +0.36 to +0.61 and +0.20 to +0.46. Cr is nearly neutral in the pooled fit (+0.024) only because its local response spans -0.15 to +0.21 and reverses sign with host ratio. Ni therefore provides the most consistent reduction benefit, whereas Cr triggers phase-conditioned competition among Mn-rich HALITE, SPINEL, and Cr-bearing oxide solutions. In multicomponent Co-Fe-Mn-Ni reduction experiments, Fe, Co, and Ni are progressively extracted while an MnO-rich skeleton persists; this supports the calculated positive Ni/Co effects and the special stability of Mn-bearing residual oxide [20,21].

*Fe-Ni host:* Mn and Cr give the strongest stabilizing responses of any additions to this host: pooled coefficients are -0.512 and -0.422, and the local ranges are -0.67 to -0.45 and -0.67 to -0.41. Co is weak ($\beta_{Co} = +0.110$) and locally fluctuates only from -0.08 to +0.04. The Fe-Ni host therefore separates the strong calculated Mn/Cr responses from the nearly neutral Co response. Environmental-TEM measurements show that Ni changes Fe-oxide reduction from a two-step to a single-step pathway and that Ni or Co can promote alloy formation and inhibit reoxidation[51]; the present calculation complements that kinetic observation by predicting that adding Mn or Cr to the Fe–Ni base moves the equilibrium onset in the opposite, oxide-stabilizing direction.

*Mn-Ni host:* Cr is weakly stabilizing on average ($\beta_{Cr} = -0.069$) and ranges from -0.27 on the Ni-rich side to approximately zero toward the Mn-rich side. Fe similarly changes from weakly negative to nearly neutral (local range -0.13 to +0.03). Co is positive in the pooled fit (+0.205) but its local curve remains close to zero (-0.08 to +0.04), showing that the pooled value is not representative of most individual host ratios. Mn-Ni therefore exhibits the greatest disagreement between edge-averaged and local interpretation. The persistent Mn-rich HALITE/MnO observed during multicomponent reduction explains why additional elements have limited leverage once Mn dominates, while Ni-rich compositions retain greater sensitivity to Cr or Fe additions [20,21,35].

Across the ten binary hosts, three robust design rules emerge : (i) Mn addition to Ni-containing binaries is consistently oxide-stabilizing and produces the largest negative coefficients; (ii) Ni addition to Mn-, Cr-, or Fe-bearing hosts is generally destabilizing, consistent with favorable NiO reduction and alloy formation, although this effect diminishes in Cr-rich local composition windows; (iii) a near-zero pooled coefficient does not imply chemical irrelevance, as demonstrated by Cr additions on Co-Mn and Fe-Mn hosts and by Co or Fe additions to Cr-Ni and Mn-Ni hosts, where substantial cancellation or phase-dependent sign changes occur.

To the best of our knowledge, no published study has systematically quantified these thirty standardized binary–third-element sensitivities. Their numerical magnitudes therefore constitute new CALPHAD predictions, while the available literature provides independent support for the underlying elemental partitioning and qualitative trends rather than direct coefficient-by-coefficient validation. This framework thus moves beyond conventional single-oxide reduction analysis by quantifying how alloying effects emerge, weaken, or reverse as a function of multicomponent composition and phase stability.

### 3.4 Stored phase boundaries and reduction complexity

HyDR of multicomponent oxides generally proceeds through multiple reduction steps rather than through a single transformative step from oxide to metal [20,49,54]. Depending on the alloying elements and temperature, the reduction pathway involves successive transitions through CORUNDUM, SPINEL, and HALITE phases before reaching the final metallic state (BCC, FCC, or HCP). Each phase transformation is associated with a distinct thermodynamic driving force for reduction. Understanding these transient phase formation steps, their interdependence, and their sequence is therefore essential for describing the overall reduction behavior of multicomponent oxide mixtures, as the sequence and stability of intermediate phases can influence the composition and the final microstructure. To gain further insights into the multistep reduction behavior, we extracted phase-boundary records and used them to characterize the thermodynamic driving force associated with each reduction step.

The stored phase-boundary records for six representative ternary compositions at 900 °C reveal substantial differences in the number of reduction boundaries and their associated energy ranges (**Figure 6**). Boundary index 1 represents the metal-side onset. At fixed temperature, the oxygen-potential equivalent derived from the phase activities should vary consistently with the oxygen chemical potential when a common reference state is used. However, some higher-index records exhibit non-monotonic energy values: the boundary energy does not increase steadily with boundary index, so that a higher-index boundary can lie below the preceding one (e.g., boundaries 2 and 3 in **Figure 6a**). Such reversals cannot be attributed to phase competition alone and therefore indicate that the stored boundary index should not be interpreted as a verified chronological or equilibrium-stability sequence. Instead, the higher-index records are treated as exploratory diagnostics that require further checking of the scan ordering, reference-state conversion and duplicate records before they can be assigned to a physical reduction sequence. The boundary-1 thresholds are considered separately, subject to the onset-search and phase-model limitations described above.

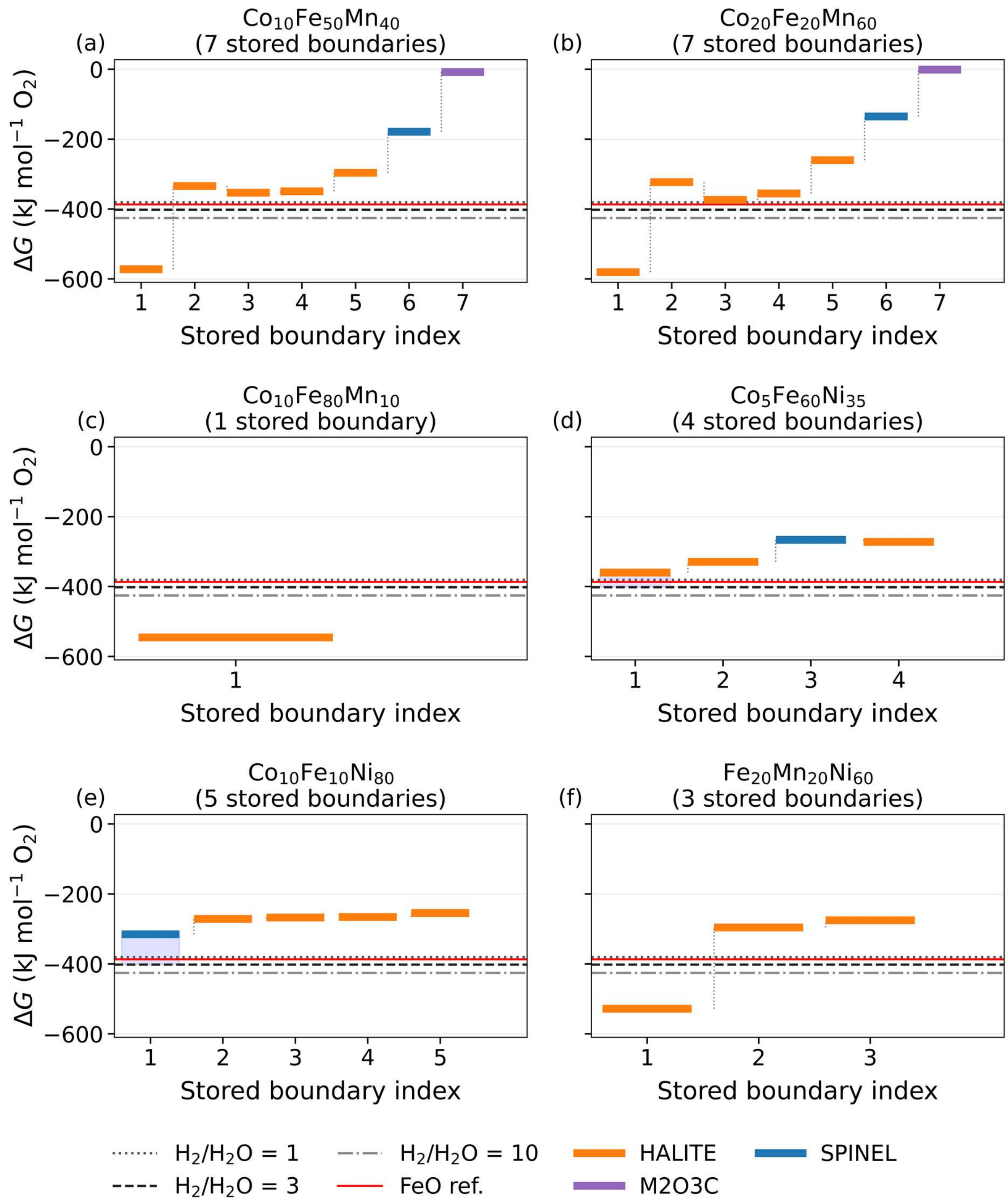

**Figure 6**: **Multi-step thermodynamic reduction pathways mixed oxides**. Multi-step reduction staircase at T = 900 °C for six representative ternary compositions. (a) $Co_{10}Fe_{50}Mn_{40}$, (b) $Co_{20}Fe_{20}Mn_{60}$, (c) $Co_{10}Fe_{80}Mn_{10}$, (d) $Co_5Fe_{60}Ni_{35}$, (e) $Co_{10}Fe_{10}Ni_{80}$ and (f) $Fe_{20}Mn_{20}Ni_{60}$. Each horizontal bar represents one sequential oxide-phase boundary (ΔG for that reduction step); bars are colour-coded by oxide phase. Vertical connectors (dotted) link successive steps, illustrating the energetic gap between them. HyDR operating lines at $H_2/H_2O = 1$(dotted), 3(dashed), and 10(dash-dot) are superimposed at -380.4, -401.8, and -425.3 kJ/mol $O_2$, respectively; the FeO reference at -386.8 kJ/mol $O_2$ is red. Blue shading marks the positive driving-force margin for boundary 1 at $H_2/H_2O = 3$, where present. The horizontal axis follows stored boundary index. Higher-index energies include non-monotonic records and require audit before the plot can be interpreted as an ordered equilibrium sequence.

The horizontal bar in **Figure 6** indicates the boundary index, its height represents equilibrium $\Delta G$, and its color identifies the oxide phase on the oxidized side of that boundary. For a given $H_2/H_2O$ ratio, a boundary is accessible when the corresponding $H_2/H_2O$ line lies below the horizontal bar because $\Delta G_{rf} = \Delta G_{boundary} - \Delta G_{HyDR} > 0$. A large vertical separation between adjacent bars represents a broad oxygen-potential interval over which an intermediate assemblage can remain thermodynamically stable, whereas closely spaced bars can be crossed within a narrow range gas composition range and may therefore appear as a single mass-loss or diffraction event. At $900°C$, increasing $H_2/H_2O$ from 1 to 10 lowers the gas line by only 44.9 $kJ/mol_{O_2}$. This shift is sufficient to change the accessibility of boundaries that lie within this energy window, but it cannot overcome a bottleneck hundreds of kilojoules below the higher-index steps.

The following composition-by-composition descriptions report the saved numerical values. Statements about accessibility compare each value with the gas line; they do not establish the validity or order of higher-index transitions. Missing records must not be interpreted as proof that no intermediate phase exists.

$Co_{10}Fe_{50}Mn_{40}$: **Figure 6a** contains seven boundaries spanning $564.5 kJ/mol_{O_2}$. Boundary 1 corresponds to HALITE at -572.2 $kJ/mol_{O_2}$, giving $\Delta G_{rf} = -170.4 kJ/mol_{O_2}$ at $H_2/H_2O = 3$. This HALITE boundary remains below even the $H_2/H_2O$ = 10 gas line, so complete metallization requires a substantially drier gas, with $r_{min} \approx 1.87 \times 10^4$. Boundary 2 jumps upward by $238.0 kJ/mol_{O_2}$ to $-334.2 kJ/mol_{O_2}$.. The next three HALITE boundaries occur at -353.2, -349.3, and $-296.5 kJ/mol_{O_2}$, followed by SPINEL at -178.8 and an $M_2O_3$-type boundary at $-7.7 kJmol_{O_2}$. Every boundary except boundary 1 lies above the $H_2/H_2O$ = 1 gas line. The calculation therefore predicts extensive partial reduction through the $M_2O_3$-SPINEL, and higher-potential HALITE fields, followed by retention of the final Mn-rich HALITE unless a much drier gas (higher $H_2/H_2O$ ratio) is used.

$Co_{20}Fe_{20}Mn_{60}$: **Figure 6b** also contains seven boundaries but has an even wider free energy variation (579.3 $kJ/mol_{O_2}$). Its HALITE bottleneck occurs at - $580.7 kJ/mol_{O_2}$ ($\Delta G_{rf} = -178.9 kJ/mol_{O_2}$ at $r = 3$; $r_{min} \approx 2.88 \times 10^4$). Boundary 2 is separated from this bottleneck by $257.8 kJ/mol_{O_2}$, while the intermediate HALITE boundaries lie between -373.7 and -260.3$kJ/mol_{O_2}$. SPINEL and $M_2O_3$-type boundaries occur at -135.2 and -1.4$kJ/mol_{O_2}$. As in $Co_{10}Fe_{50}Mn_{40}$, all six higher-index boundaries are accessible even at $r = 1$, whereas complete reduction remains thermodynamically inaccessible because of the isolated low-energy HALITE onset. Increasing the Mn content therefore affects not only the number of reduction steps but also the depth of the final bottleneck and the oxygen-potential range over which intermediate oxide phases can remain thermodynamically stable.

$Co_{10}Fe_{80}Mn_{10}$: **Figure 6c** shows only one HALITE boundary, at $-545.5\,\mathrm{kJ/mol_{O_2}}$. The absence of higher-index boundaries indicates that the equilibrium scan predicts no stable intermediate reduction staircase at $900°C$; it does not, however, imply that reduction is thermodynamically feasible. The single boundary remains below all three $H_2/H_2O$ lines, with $\Delta G_{rf} = -143.7\mathrm{kJ/mol_{O_2}}$ at $r = 3$ and $r_{min} \approx 4.75 \times 10^3$. This demonstrates that the number of predicted reduction steps (transient phases) is distinct from the thermodynamic difficulty of reduction: a chemically one-step pathway can still have a larger thermodynamic driving force for reduction.

$Co_5Fe_{60}Ni_{35}$: **Figure 6d** shows four boundaries occupying a narrow interval of $93.6\mathrm{kJ/mol_{O_2}}$. Boundary 1 is HALITE at $-360.2\mathrm{kJ/mol_{O_2}}$, followed by HALITE at $-329.1\,\mathrm{kJ/mol_{O_2}}$, SPINEL at $-266.6\mathrm{kJ/mol_{O_2}}$, and a final HALITE boundary at $-272.3\,\mathrm{kJ/mol_{O_2}}$. The last two boundaries are inverted by 5.8 $kJ/mol_{O_2}$ (the final HALITE boundary lies slightly below the preceding SPINEL boundary), so this part of the profile should not be read as an ordered equilibrium sequence. All four boundaries lie above the $H_2/H_2O = 1$ gas line; the boundary-1 margin is already +20.2 $kJ/mol_{O_2}$ at $r = 1$ and +41.6 $kJ/mol_{O_2}$ at $r = 3$. Thus, both the intermediate transformations and complete reduction are thermodynamically accessible throughout the displayed $H_2/H_2O$ window.

$Co_{10}Fe_{10}Ni_{80}$: **Figure 6e** contains five closely grouped boundaries across only 60.7 $kJ/mol_{O_2}$. The controlling SPINEL boundary is -315.4 $kJ/mol_{O_2}$, and four HALITE boundaries cluster from -271.6 to -254.6 $kJ/mol_{O_2}$; three of those are separated by only 1-12 $kJ/mol_{O_2}$. Every step is accessible at $r = 1$, and the boundary-1 driving force increases from +65.0 $kJ/mol_{O_2}$ at $r = 1$ to +86.4 $kJ/mol_{O_2}$ at $r = 3$. The tightly packed HALITE transitions may consequently appear as a single broadened event, whereas the SPINEL-to-metal approach should remain thermodynamically favourable.

$Fe_{20}Mn_{20}Ni_{60}$: **Figure 6f** contains three HALITE boundaries at -528.5, -296.0, and -275.3 $kJ/mol_{O_2}$. The 232.5 $kJ/mol_{O_2}$ gap between boundaries 1 and 2 divides the pathway into two regimes: the two higher-index transformations are accessible even at $r = 1$, but the final

HALITE boundary remains inaccessible at $r = 10$ . At $r = 3$ its driving force is -126.7 $kJ/mol_{O_2}$and the calculated threshold is $r_{min} \approx 1.98 \times 10^3$. The presence of substantial Ni destabilizes the oxide mixtures but does not provide enough driving force to overcome HALITE bottleneck associated with MnO formation.

Overall, **Figures 6a-f** demonstrate that the complexity of the intermediate reduction pathway is distinct from the thermodynamic criterion (driving force) for complete metallization. The Mn-containing Co–Fe–Mn and Fe–Mn–Ni compositions undergo most of their intermediate reduction boundaries under the displayed atmospheres, yet remain limited by the boundary-1 HALITE. In contrast, both Co–Fe–Ni compositions can cross all predicted boundaries even at $r = 1$. Notably, Fe-rich one-step composition ($Co_{10}Fe_{80}Mn_{10}$) exhibits a substantially larger thermodynamic driving-force requirement for reduction than the more complex four- and five-step Co–Fe–Ni pathways. Thus, a greater number of intermediate transformations does not necessarily imply a larger thermodynamic requirement for complete metallization.

Boundary 1 (the metal-side onset boundary) in **Figure 6** is analogous to the classical wüstite-to-Fe bottleneck in iron-oxide reduction. Above the wüstite stability temperature (>570 °C), $Fe_2O_3$ reduces through $Fe_3O_4$ and FeO before metallic Fe forms, with the final monoxide-to-metal transformation often governing the late-stage thermodynamic and kinetic response [5,11,55]. Below 570 °C, wüstite is unstable and $Fe_3O_4$ reduces directly to Fe, so the final step is the magnetite-to-metal transformation. The present calculations span both regimes: the 500 and 550 °C results lie in the low-temperature regime and the 600–1500 °C results above the wüstite stability limit. In the mixed oxides, however, a HALITE solid solution can remain the last oxide even below 570 °C, because MnO, CoO and NiO stabilize the rock-salt structure (Figure 3 shows HALITE onsets down to 500 °C). The HALITE boundary identified here should not be compared with pure FeO, as HALITE represents a multicomponent (Fe, Mn, Co, Ni)O solid solution. Nevertheless, its role as the final persistent monoxide field is thermodynamically analogous to that of FeO in conventional iron-oxide reduction.

The wide Mn-rich staircases in **Figures 6a**, **b** and **f** are consistent in-situ HyDR studies conducted on a Co-Fe-Mn-Ni oxide mixture, which identified SPINEL, HALITE, and $Mn_3O_4$ as transient phases, followed by the formation of Fe-Co-Ni alloy while MnO remained as the unreduced oxide [20,21]. These studies further revealed different reduction pathways for mechanically mixed and pre-sintered precursors, despite achieving similar reduction degrees (~80%). Accordingly, the persistence of boundary-1 HALITE provides the strongest point of comparison with the present calculations. A one-to-one correspondence between every higher-index boundary and an experimentally observed phase is not expected, because precursor history and kinetic effects can cause intermediate phases to merge, bypass, or remain unresolved during reduction.

The narrower, fully accessible Co-Fe-Ni profiles shown in **Figure 6d** and **e** are consistent with experimental evidence that Ni alters Fe-oxide reduction pathways and lowers the overall reduction activation energy by 18.79 % [51,56]. HyDR of $Fe_2O_3$ + NiO oxide mixtures further report that early Ni metallization accelerates Fe-oxide reduction through hydrogen spillover, thereby lowering the reduction onset temperature [57]. These kinetic mechanisms are not contained in the CALPHAD staircase but support the composition-dependent reduction trend predicted here. The superimposed $H_2/H_2O$ gas lines also have experimental relevance: water addition modifies the local reduction potential and slows NiO reduction kinetics [46,58], supporting $H_2/H_2O$ as a suitable control variable while highlighting that thermodynamic accessibility of a boundary does not determine the time required to complete the corresponding reduction step.

The residual oxide at the metal-side threshold defines the equilibrium endpoint of reduction. However, the higher-index boundary records should not be interpreted as a one-to-one prediction of transformation order or experimental mass-loss events. Establishing such a correspondence would require phase fractions, oxygen stoichiometry to relate equilibrium boundaries to the thermogravimetric reduction results. The staircase should therefore be regarded as a diagnostic representation of the calculated thermodynamic boundaries rather than a validated reaction pathway. Chemo-mechanical phase-field modeling of iron-oxide reduction explicitly couples phase transformation, species transport and deformation as the additional physical description required to predict evolving morphology and reaction kinetics [59], which builds on the diffuse-interface descriptions of Cahn and Hilliard and of Allen and Cahn [60,61]. The present boundary records contain no equivalent information. Nevertheless, they provide a useful thermodynamic basis for identifying which intermediate oxide states can persist and, most importantly, which residual oxide ultimately limits complete metallization.

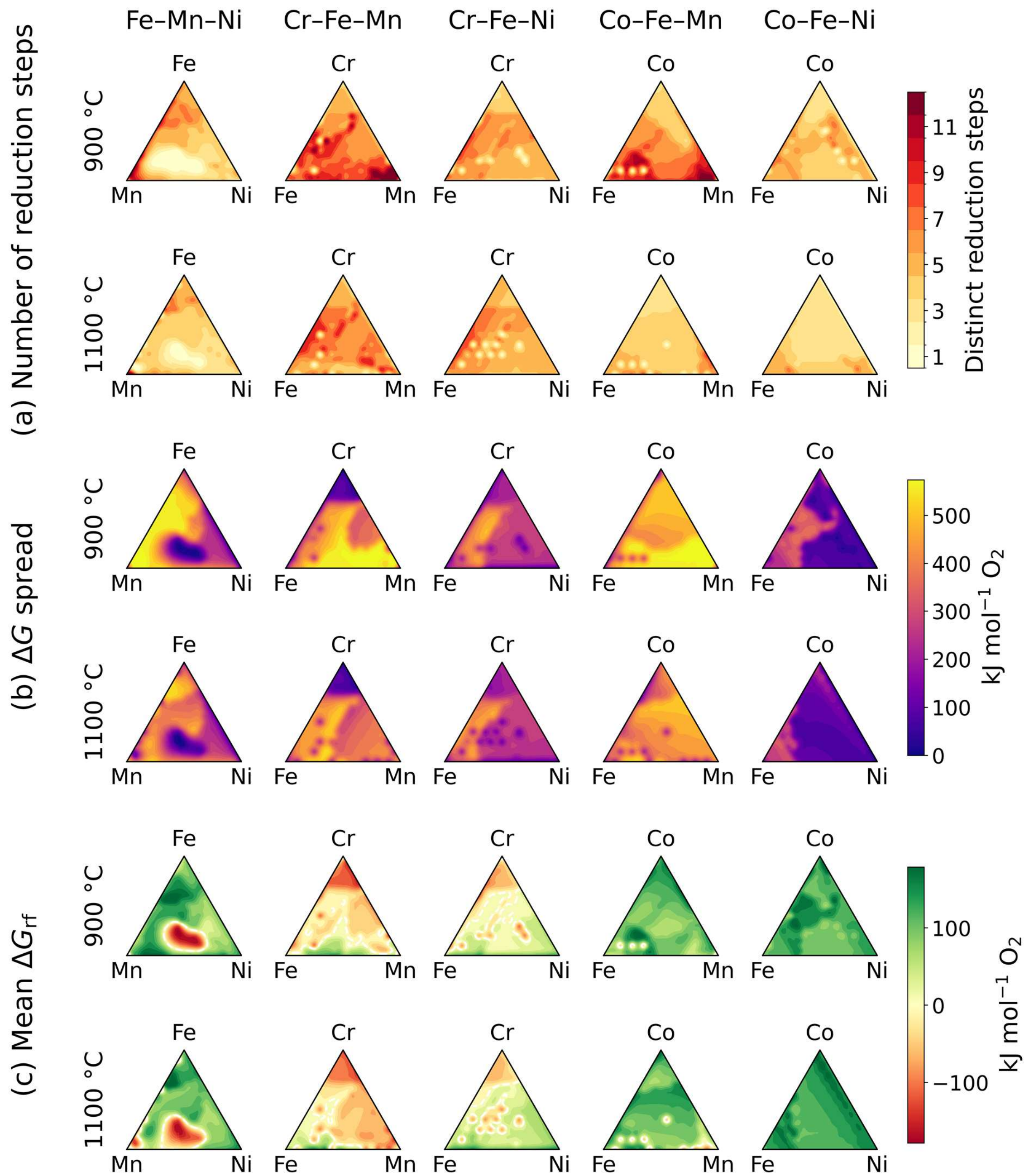

**Figure 7**: **Mixed-oxide thermodynamic stability and reduction window for ternary sub-systems**. Mixed-oxide reduction complexity across five Fe-bearing ternary domains at 900 and 1100 °C. Columns correspond to the five ternary systems. Each pair of rows (900 and 1100 °C) uses a common colour scale for (a) the number of candidate boundaries, (b) the ΔG spread of the recorded boundary energies and (c) the unweighted mean reduction driving force at $H_2/H_2O = 3$. White dashed contours in (c) mark zero mean driving force, not the final metal/oxide threshold. Heating contracts the high-count and broad-spread regions most clearly in Co–Fe–Mn and Co–Fe–Ni; the mean-driving-force response varies with composition. The maps describe calculated boundary records and do not specify kinetic reaction steps or reduction times.

The three paired-row groups in **Figure 7** represent three descriptors of the calculated reduction behavior: (a) the number of candidate boundaries, (b) the spread of their associated energies, and (c) their position relative to the imposed $H_2/H_2O$ ratios. For a composition with N stored boundaries, where $\Delta G_j$ is the ΔG of boundary j (j = 1, …, N), the ΔG spread plotted in **Figure 7b** is defined as $\Delta G_{spread} = \max_j(\Delta G_j) - \min_j(\Delta G_j)$, expressed in kJ $mol^{-1}$ $O_2$. A single boundary has therefore zero spread irrespective of whether that boundary is readily accessible or requires a high reduction potential. The plotted step-count metric describes candidate transitions stored in the equilibrium calculation and should not be interpreted as the number of experimentally resolved reactions; neither the label "distinct reduction steps" nor interpolation establishes chronological ordering. The mean driving force is the arithmetic mean of $\Delta G_j - \Delta G_{HyDR}$ over the boundary records included in each composition. It is not weighted by phase fraction or oxygen removal. Consequently, a positive mean value can coexist with a negative driving force at boundary 1, which represents the final oxide-removal threshold in the reverse equilibrium direction.

**Figure 7a** shows that temperature-induced changes in candidate-boundary count are strongest in the Co-containing domains. In Co–Fe–Mn, the high-count regions near the Fe–Mn edge at 900 °C contract markedly at 1100 °C, while Co–Fe–Ni becomes dominated by low-count compositions. In contrast, Cr–Fe–Mn retains a heterogeneous band of higher counts, and Cr–Fe–Ni maintains appreciable intermediate complexity. These contrasts indicate that increasing temperature (900 to 1100 °C) changes the number of candidate phase-field crossings in a composition-dependent manner. Importantly, they do not establish that a high-count composition necessarily has the most stable residual oxide: one persistent boundary can impose a higher $H_2/H_2O$ requirement than several weakly separated boundaries.

**Figure 7b** reveals a complementary aspect of these reduction landscapes. Broad energy spreads occur over substantial portions of the Mn-bearing domains, particularly Co–Fe–Mn and Cr–Fe–Mn, whereas Co–Fe–Ni exhibits a much narrower spread at 1100 °C. Within Fe–Mn–Ni, a low-spread region near the Mn–Ni side contrasts with the broader-spread region extending towards the Fe–Mn side. The corresponding mean-driving-force map contains a negative region, showing that closely clustered boundaries can all lie on the unfavorable side of the imposed gas line. Conversely, a wide energy spread can include both readily crossed intermediate boundaries and a strongly retained final oxide. Thus, spread therefore measures the separation of stored stability thresholds, rather than an activation energy, heat requirement or number of oxygen atoms removed.

**Figure 7c** further illustrates this distinction in Co–Fe–Mn. A substantial fraction of this composition space exhibits a positive mean driving force at both temperatures, although the boundary-1 maps identify extensive regions where complete oxide removal remains unfavorable at $H_2/H_2O$ = 3. Favorable intermediate boundaries therefore increases the mean without eliminating the Mn-associated endpoint constraint. Fe–Mn–Ni similarly contains a localized negative-mean region within a predominantly positive-mean field. Cr–Fe–Mn and Cr–Fe–Ni retain unfavorable regions toward the Cr-rich apex, whereas Co–Fe–Ni remains on the favorable side of the mean-driving-force criterion. Thus, the zero contours therefore identify the compositions where the average boundary energy coincides with the imposed gas atmosphere; they should not be interpreted as contours of complete metallization.

### 3.5 Comparison with published literature and computational scope

The Cr-bearing compositions provide a complementary comparison with the work of Al-Buainain and Dunand [47], who demonstrated the sequential hydrogen reduction of $Fe_2O_3$–NiO–$Cr_2O_3$ oxide blends towards Fe–Cr–Ni stainless steels during heating to 1300 °C, with residual $Cr_2O_3$ (< 1.1 vol. %) retained after HyDR completion. Their observations are consistent with alloy-assisted reduction in Fe/Ni-containing blends and with the persistence of a thermodynamically stable Cr-rich residual oxide. This experimental evidence supports the present thermodynamic prediction that Cr-bearing compositions can retain a stable oxide at the final reduction boundary. However, the oxide compositions, reduction temperatures, reducing atmosphere, and precursor history differ from those used in the present calculations. Therefore, the comparison is qualitative rather than a direct point-by-point validation of the calculated mean driving force or residual oxide fraction.

The comparatively simple Co–Fe–Ni landscape provides another experimental comparison. Pan et al. reported that Ni incorporation changed iron-oxide reduction from a two-step ($Fe_3O_4$ → FeO → Fe) to a single-step ($Fe_3O_4$ → Fe) mechanism using environmental transmission electron microscopy [51] . This observation supports the broader conclusion that alloying can alter intermediate-phase evolution rather than simply shifting the reduction sequence of the individual oxides. The lower boundary counts and narrower energy spreads predicted for Co–Fe–Ni therefore provide an equilibrium counterpart to the experimentally observed simplification of the reduction pathway.

Three observations emerge from the combined modelling and experimental comparisons. First, the hydrogen requirement is strongly composition dependent: the sampled Co–Fe–Ni compositions are fully reducible at $H_2/H_2O = 3$, whereas Mn- and Cr-containing composition spaces retain broad regions with unfavorable driving forces and therefore require higher $H_2/H_2O$ ratios. Second, this behavior is structural rather than a simple elemental ordering. Mn-rich HALITE and Cr rich CORUNDUM set the final reduction boundary, while multicomponent SPINEL introduces additional oxygen potentials thresholds. Third, increasing temperature generally broadens the thermodynamic reduction margins, although the relative response of individual composition domains can change. These results highlight that precursor composition and gas humidity ($H_2/H_2O$ ratio) are coupled but independent design variables for achieving complete metallization.

The calculated elemental partitioning among oxide phases also agrees with independent oxidation studies of Cantor-type alloys. Experimental studies have reported Mn-rich outer oxides, Cr-rich inner CORUNDUM, and enrichment of Ni, Co, and Fe beneath the oxide scale [28,35,50,62]. These observations provide experimental evidence for the preferential partitioning of alloying elements among oxide and metallic regions, consistent with the composition-dependent phase stability captured by the thermodynamic framework in this study. The sensitivity analysis (**Figure 4** and **5**; **Supporting Table S3**; **Figure S11** and **S12**) further resolves this composition dependence: Mn and Cr dominate the pooled and local responses, while pairwise terms capture deviations associated with phase selection and gradient boosting models describe residual curvature near phase-field transitions. Composition-blocked validation shows that these patterns interpolate the CALPHAD surface for compositions not used during fitting. Importantly, this separation between calculation fidelity and experimental validity prevents the statistical scores from being presented as experimental agreement.

The framework deliberately does not predict reduction time, reduction rate or scale morphology. Nucleation, diffusion, porosity, segregation, precursor history, gas utilization and the difference between inlet and interfacial humidity can alter reduction rates and transient phase sequences [5,57,63]. These effects may alter the local conditions and time required to reach the equilibrium, but do not alter the equilibrium boundary at a given thermodynamic state. Thus, the principal quantitative claim of the present framework is the composition-dependent final-boundary threshold and associated $H_2/H_2O$ requirement for complete metallization.

## 4 Conclusions

The thermodynamic framework developed in this study establishes residual-oxide stability as the central equilibrium criterion for designing multicomponent alloys via hydrogen-based direct reduction (HyDR) of Co–Cr–Fe–Mn–Ni mixed oxides. Linking the final metal–oxide boundary to the $H_2/H_2O$ ratio provides a common basis for comparing precursor compositions and desired reduction atmosphere required for complete metallization. Co–Fe–Ni mixtures exhibit comparatively favorable reduction windows, whereas Mn-rich HALITE and Cr-rich CORUNDUM impose persistent residual-oxide boundaries across broad composition regimes. Consequently, extensive reduction of other constituent oxides can occur while a Mn- or Cr-rich oxide remains stable.

Alloying effects depend strongly on the surrounding composition: although Mn and Cr generally increase the stability of the oxide mixtures (e.g., adding 10 at.% Cr or Mn to an equiatomic Fe–Ni oxide mixture lowers the onset free energy at 900 °C by about 168 and 161 kJ $mol^{-1}$ $O_2$, respectively, and raises the $H_2/H_2O$ ratio required for complete reduction from about 0.3 to above $10^3$, whereas 10 at.% Co changes it by less than 1 kJ $mol^{-1}$ $O_2$), the magnitude and, in some composition ranges, the direction of an elemental effect can change with the controlling oxide phase. For example, Cr stabilizes Co–Mn and Fe–Mn oxide mixtures at some host ratios but destabilizes them at others. The local sensitivity and interaction analyses therefore show why pure-oxide rankings or a single average elemental coefficient are insufficient to describe concentrated multicomponent mixtures. Nonlinear statistical models help interpolate these composition-dependent trends, but their predictive performance measures agreement with the calculated thermodynamic surface, rather than independent validation of the underlying thermodynamic description.

The analysis further demonstrates that intermediate phase complexity and the thermodynamic boundary condition for complete metallization are distinct. A small number of stored boundaries does not imply that the reduction is thermodynamically feasible, while a favorable average driving force does not guarantee reduction of the final residual-oxide. The composition-specific residual-oxide threshold remains the primary endpoint criterion, with higher-index boundary counts and energy spreads serving as complementary descriptors of the intermediate equilibrium landscape rather than validated reaction sequences. These findings provide a thermodynamic basis for jointly selecting alloy composition, oxide precursor chemistry and reduction conditions for sustainable alloy design from mixed oxide precursors.

## 5 Data availability

The computational data (CSV files), TC-Python scripts, analysis and figure-generation code, and reproduction instructions are available in a public GitHub repository at https://github.com/albert-hzbn/CoCrFeMnNi-hydrogen-reduction. Reanalysis of the archived data requires no Thermo-Calc license; regenerating the thermodynamic calculations requires licensed Thermo-Calc and TCOX10 software.

## 6 CRediT authorship contribution statement

**A. L.** Conceptualization, Methodology, Investigation, Formal analysis, Validation, Data Curation, Visualization, Writing – original draft; **S. S.** Conceptualization, Methodology, Visualization, Writing – review & editing; **D. R.** Supervision, Funding acquisition, Resources, Writing – review & editing.

## 7 Declaration of Competing interests

The authors declare that they have no known competing financial interest or personal relationships that could have appeared to influence the work reported in this paper.

## 8 Acknowledgments

A.L. acknowledges support from the Max Net Initiative SusMax – The Max Planck Network on Sustainable Materials. S.S. acknowledges the financial support from Horizon Europe project HAlMan co-funded by the European Union grant agreement (ID 101091936). D.R. is grateful for financial support from the European Union through the ERC Advanced grant ROC (Grant Agreement No. 101054368). Views and opinions expressed are however those of the author(s) only and do not necessarily reflect those of the European Union or the ERC. Neither the European Union nor the granting authority can be held responsible for them.

## 9 Supplementary material

Supplementary information is available for this paper.

## 10 Declaration of generative AI and AI-assisted technologies in manuscript preparation

OpenAI Codex was used to assist manuscript restructuring, literature checks, consistency review and scripted document editing. AI-assisted outputs do not constitute experimental evidence. The authors retain responsibility for verifying the scientific content, references, calculations and final submission.

# Supplementary Material

## Designing sustainable high-entropy alloys directly from mixed oxides: A thermodynamic perspective

Albert Linda*[,1], Shiv Shankar*[,1], Dierk Raabe

Max Planck Institute for Sustainable Materials, Max-Planck-Str. 1, 40237, Düsseldorf, Germany

*Corresponding author: a.linda@mpi-susmat.de; s.shankar@mpi-susmat.de

[1] These authors contributed equally to this work

This Supporting Information documents the computational populations, thermodynamic extraction, map construction and statistical diagnostics underlying the manuscript. **Figures S1–S10** compare oxidation-onset energies and phase identity; **Figure S11** assesses a statistical model fitted to those calculations. **Figure S12** reports the temperature-resolved standardized coefficients $\beta_k$ discussed in the main text. **Tables S1–S4** report extraction routes and model summaries, while **Tables S5–S9** provide sampling counts, model selection, error tails, composition-grouped uncertainty and file provenance. All results are computational; no new experiments are reported.

## Definitions and interpretation of the figures

In **Figures S1–S10**, n is the number of stored oxidation-onset calculation records contributing to that panel at the stated temperature and within the stated ternary face. It is not the number of experimental specimens, replicates, interpolated pixels or unique compositions. **Figures S1–S5** include all boundary-index-1 records with finite onset energy; **Figures S6–S10** additionally require a valid phase label. Repeated records at the same composition and temperature remain in n. **Table S5** reports n alongside the number of distinct metal compositions, so the sampling density is explicit.

A ternary face contains compositions formed from the three named metals, including its binary edges and pure-element vertices; the other two metal contents are zero. Atomic percentages are normalized to the metal total and sum to 100%. The top, left and right vertices correspond to the printed element labels. Panel order is (a) 700, (b) 900, (c) 1100 and (d) 1300 °C. Differences in n across temperature reflect the coverage of the assembled calculation file and, for phase maps, label screening; they are not evidence of a temperature-dependent number of physical specimens.

Boundary index 1 denotes the first oxide boundary reached while increasing oxygen chemical potential from the metallic side. In reverse equilibrium reduction, it represents the last oxide-bearing boundary before the modeled metal-only state. Higher indices identify stored candidate boundaries; index 0 is a no-boundary or failure flag. A valid phase label is nonempty and every token separated by "|" contains at least two characters. For coexisting labels, the first-listed phase is used for the categorical maps. This convention does not establish that this phase has the largest amount.

Thermodynamic calculations use Thermo-Calc 2025a and TCOX10 at 1 bar with the homogeneous FCC metallic-phase restriction described in the manuscript. The assembled file contains 310,749 records and 3,526 unique metal compositions. Of 55,051 onset records, 54,385 pass the phase-label screen and 35,934 additionally contain Fe. There are 2,973 repeated composition–temperature onset keys beyond the first occurrence and 1,399 valid onset records with coexisting phase labels. These populations serve different analyses and must not be interchanged.

In **Figures S1–S5**, the white dashed contour denotes $H_2/H_2O = 3$ and the cyan dotted contour $H_2/H_2O = 1$, each outlined in black for contrast. At a fixed composition, a positive difference between onset energy and gas oxygen-potential equivalent indicates equilibrium feasibility; zero denotes the boundary. The ratios refer to local partial pressures at the reacting surface. A colored region or phase label does not specify reaction rate, phase amount or complete reduction under a particular inlet-flow condition.

## Map construction and numerical meaning

Continuous onset maps use metal fractions as barycentric coordinates, with horizontal coordinate equal to the right fraction plus half the top fraction and vertical coordinate equal to $\sqrt{3}/2$ times the top fraction. Repeated coordinates are median-aggregated for energy interpolation. A local thin-plate radial-basis interpolator uses up to 64 neighboring coordinates, polynomial degree 1 and smoothing equal to $10^{-8}$ times the square of the larger of the energy range or unity. The field is evaluated on a 200 × 200 grid and clipped to the sampled energy range. Linear interpolation with nearest-neighbor filling is the fallback. Completion outside the sampled convex hull remains an extrapolated visual estimate within the ternary triangle.

For each ternary energy figure, the color limits are the 2nd and 98th percentiles across that face's stored onset records over the available temperatures, with 20 color intervals and extensions for values outside the displayed limits. All four temperature panels of a figure share this scale, but different ternary figures need not. Percentages quoted in captions count stored records satisfying the gas criterion before interpolation. Because repeated records contribute to these percentages, they are neither area fractions of the triangle nor uniformly weighted fractions of unique compositions.

Categorical phase maps aggregate repeated coordinates into phase-membership fractions. Gaussian-kernel interpolation uses a bandwidth equal to the larger of 0.025 or 0.60 times the median nearest-coordinate distance in barycentric Cartesian coordinates. Memberships are evaluated on a 500 × 500 grid; the class with the largest interpolated membership is displayed. Thus, narrow fields and smooth boundaries are visual estimates whose resolution is limited by the original composition sampling. The plots use 600 dpi raster images for readability; image resolution does not increase thermodynamic sampling resolution.

The oxygen-potential equivalent is reported per mole of $O_2$. Temperature in thermodynamic expressions is in kelvin, whereas panel labels use degrees Celsius. The gas expression is $[-483600 + 88T - 2RT \ln(r)]/1000$ in kJ mol$^{-1}$ $O_2$, with $R = 8.314$ J mol$^{-1}$ K$^{-1}$ and $r = p(H_2)/p(H_2O)$. The positive driving-force convention subtracts this gas value from the onset energy. The linear gas expression is a screening approximation. An illustrative ±10 kJ mol$^{-1}$ $O_2$ energy perturbation changes the equilibrium gas ratio by a multiplicative factor of approximately 1.55 at 1100 °C; this is a sensitivity illustration, not a quantified database uncertainty.

## Thermodynamic extraction and fitted phase lines

***Table S1:*** *Thermodynamic extraction routes combined in the final dataset.*

| Route | Detection principle | Domain and resolution | Purpose |
|---|---|---|---|
| Adaptive phase-boundary scan | Changes in the stable phase assemblage with adaptive boundary refinement | $\mu(\mathrm{O}) = -600$ to $-100\,\mathrm{kJ\,mol^{-1}}$ O; 200 coarse and 25 refined steps | Stored candidate phase boundaries; ordering requires verification |
| Onset-only equilibrium search | First oxygen-rich condensed phase; bracket, bisection, and direct-grid fallback | 100 scan divisions, 20 bisections, 25-point fallback grid | Recover zero-boundary sentinels |
| Face-grid boundary scan | Boundary extraction on lower-order faces | Binary and ternary faces at 5 at.%; quaternary faces at 10 at.% | Increase composition coverage |
| Extended oxygen-potential scan | Adaptive scan with a higher oxidising limit | Co-bearing binary and ternary cases; upper limit $-44\,\mathrm{kJ\,mol^{-1}}$ O | Recover high-temperature CoO onset |
| High-resolution boundary scan | Denser coarse oxygen-potential scan | Twelve Fe–Mn–Ni cases; 1000 coarse steps | Resolve narrow Mn-oxide windows |
| Label-targeted recalculation | Full equilibrium rerun after phase-label screening | Empty, one-character, or truncated phase tokens | Restore phase identity without label imputation |

***Table S2**: Recomputed pooled linear fits ΔG = A + BT for the current calculation file. Each valid boundary-index ≥ 1 record is assigned to its first-listed oxide phase; repeated records are retained. N is the number of fitted records. A is in kJ $mol^{-1}$ $O_2$, B in kJ $mol^{-1}$ $O_2$ $K^{-1}$ and T is in kelvin. These are descriptive pooled fits, not standard formation energies of pure compounds.*

| Phase | N | A | B | $R^2$ | Tmin °C | Tmax °C |
|---|---|---|---|---|---|---|
| SPINEL | 67,517 | -537.0 | +0.137 | 0.060 | 500 | 1500 |
| HALITE | 129,335 | -532.7 | +0.130 | 0.068 | 500 | 1500 |
| CORUNDUM | 58,260 | -482.8 | +0.058 | 0.010 | 500 | 1500 |
| CUB_A13 | 2,483 | -792.1 | +0.187 | 0.379 | 500 | 1150 |
| ALPHA_SPINEL | 10,067 | -283.6 | +0.144 | 0.080 | 500 | 1150 |
| M2O3C | 5,119 | -173.6 | +0.143 | 0.638 | 500 | 1150 |
| NIMNO3 | 320 | -193.5 | +0.185 | 0.686 | 500 | 800 |
| NI6MNO8_TYPE | 1,266 | -226.9 | +0.217 | 0.682 | 500 | 750 |

The fits use ordinary least squares across all compositions and temperatures assigned to each phase. A phase label identifies a solution structure with variable cation composition, not one fixed compound. Low $R^2$ therefore indicates that a single temperature line represents little of the pooled variation; it cannot be interpreted as an independently validated phase stability model. The slope B also contains composition-coverage effects and must not be treated as the entropy of a fixed reaction.

## Individual ternary onset maps

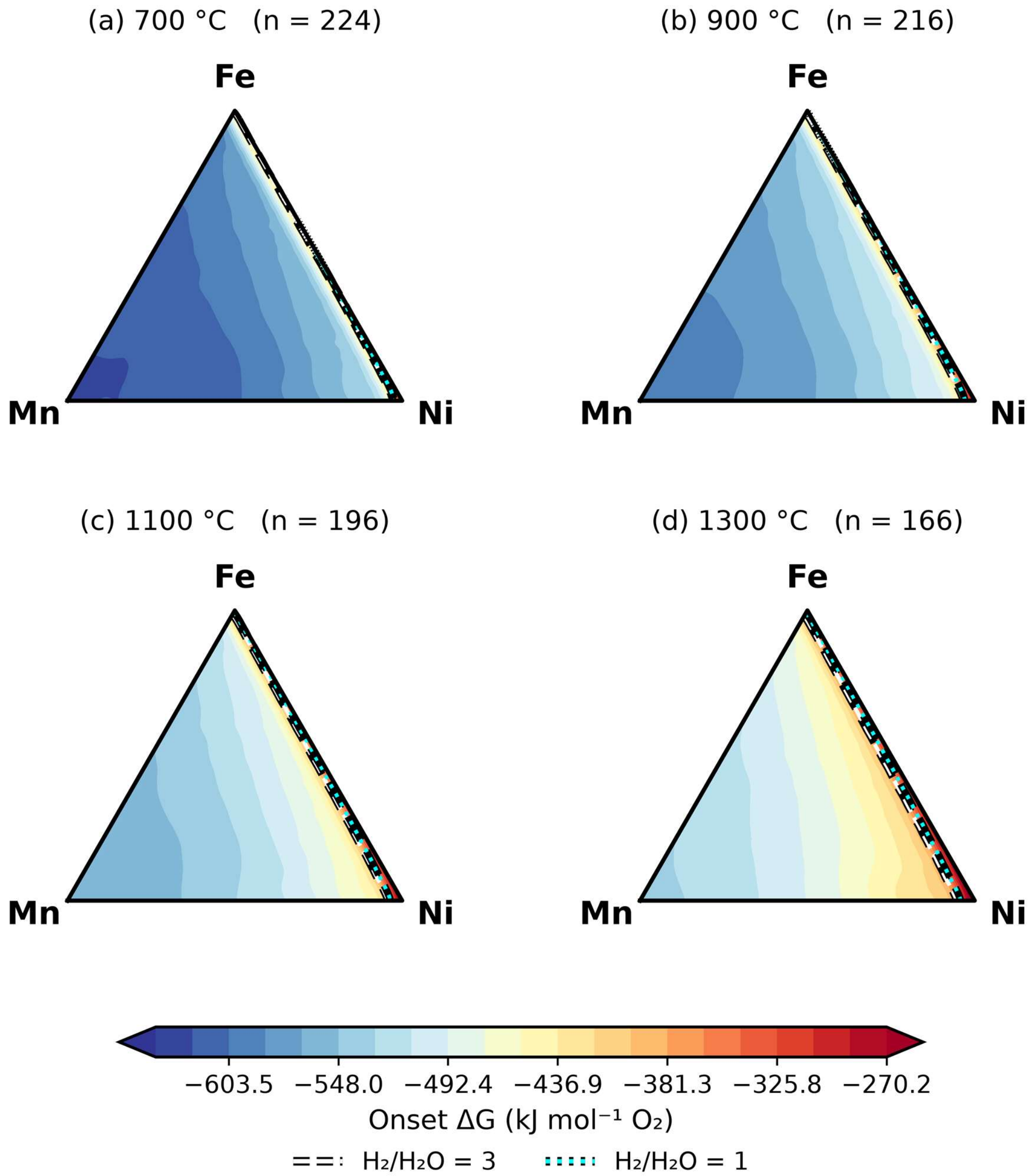

***Figure S1:*** *ΔG onset on the Fe–Mn–Ni face at four temperatures. The white contour is the $H_2/H_2O$=3 equilibrium boundary. Only 14–15% of sampled points lie on the reducible side at each displayed temperature. Here n counts stored onset calculation records at that temperature, including repeated compositions; see* ***Table S5*** *for unique-composition counts.*

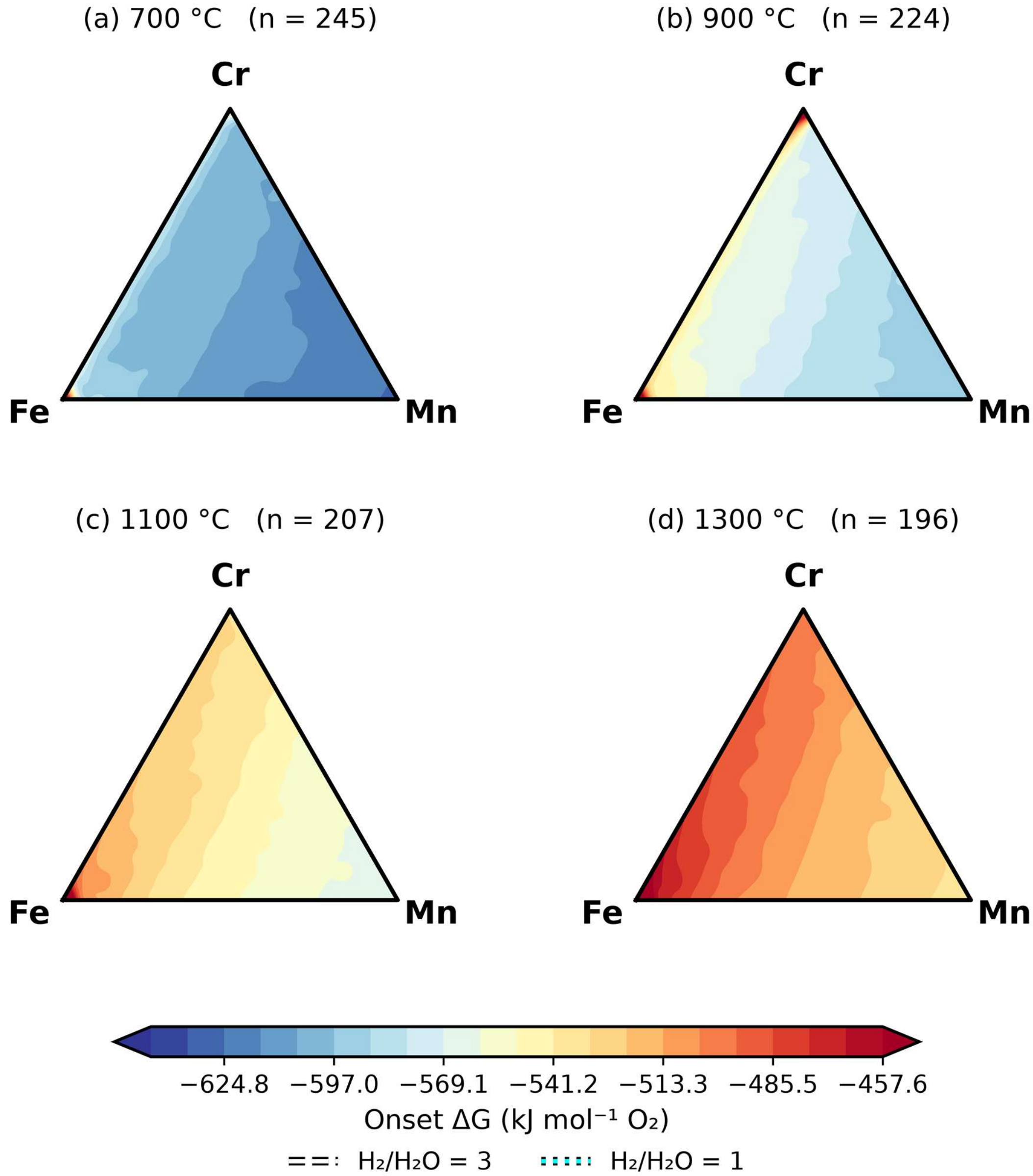

***Figure S2:*** *Onset ΔG on the Cr–Fe–Mn face. Cr-rich and Mn-rich compositions retain strongly negative onset values. The sampled reducible fraction at $H_2/H_2O$ = 3 is 0.5–1.3%. Here n counts stored onset calculation records at that temperature, including repeated compositions; see* ***Table S5*** *for unique-composition counts.*

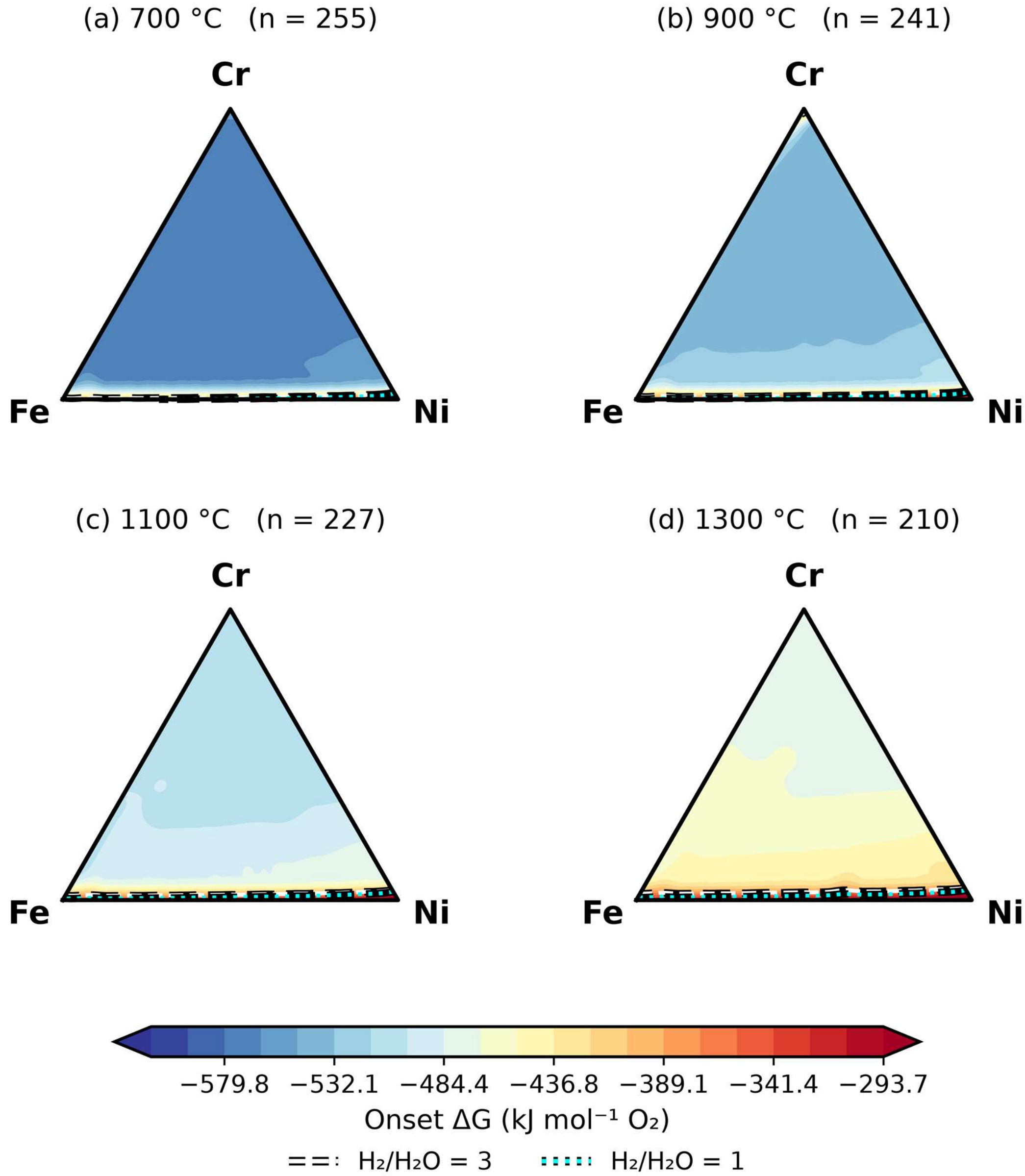

***Figure S3:*** *Onset ΔG on the Cr–Fe–Ni face. Cr produces a steep stability gradient, whereas the Ni-rich region has the largest reduction margin. Here n counts stored onset calculation records at that temperature, including repeated compositions; see* ***Table S5*** *for unique-composition counts.*

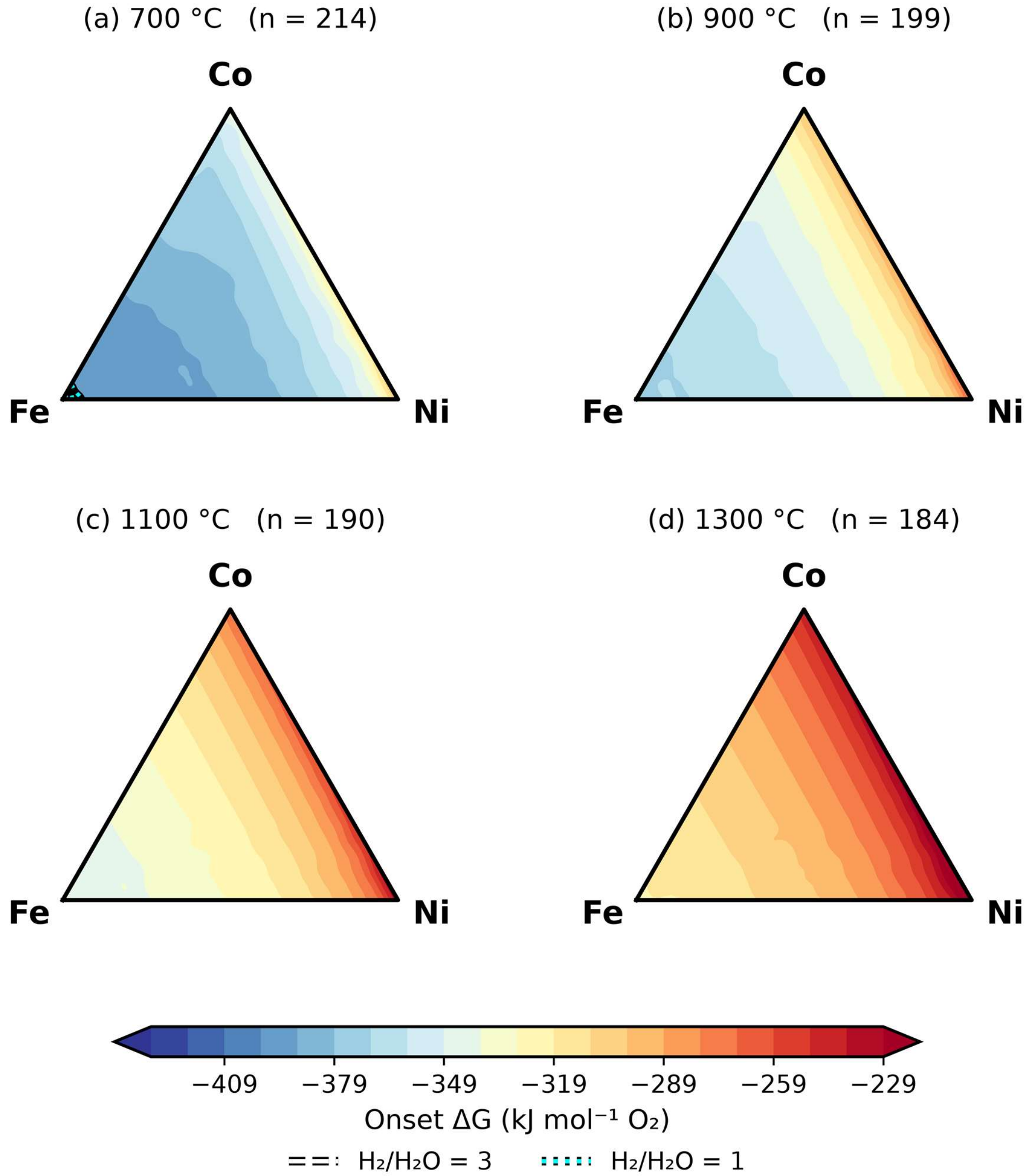

***Figure S4:*** *Onset ΔG on the Co–Fe–Ni face. Every directly sampled composition has positive driving force at $H_2/H_2O$ = 3 across the four displayed temperatures. Here n counts stored onset calculation records at that temperature, including repeated compositions; see* ***Table S5*** *for unique-composition counts.*

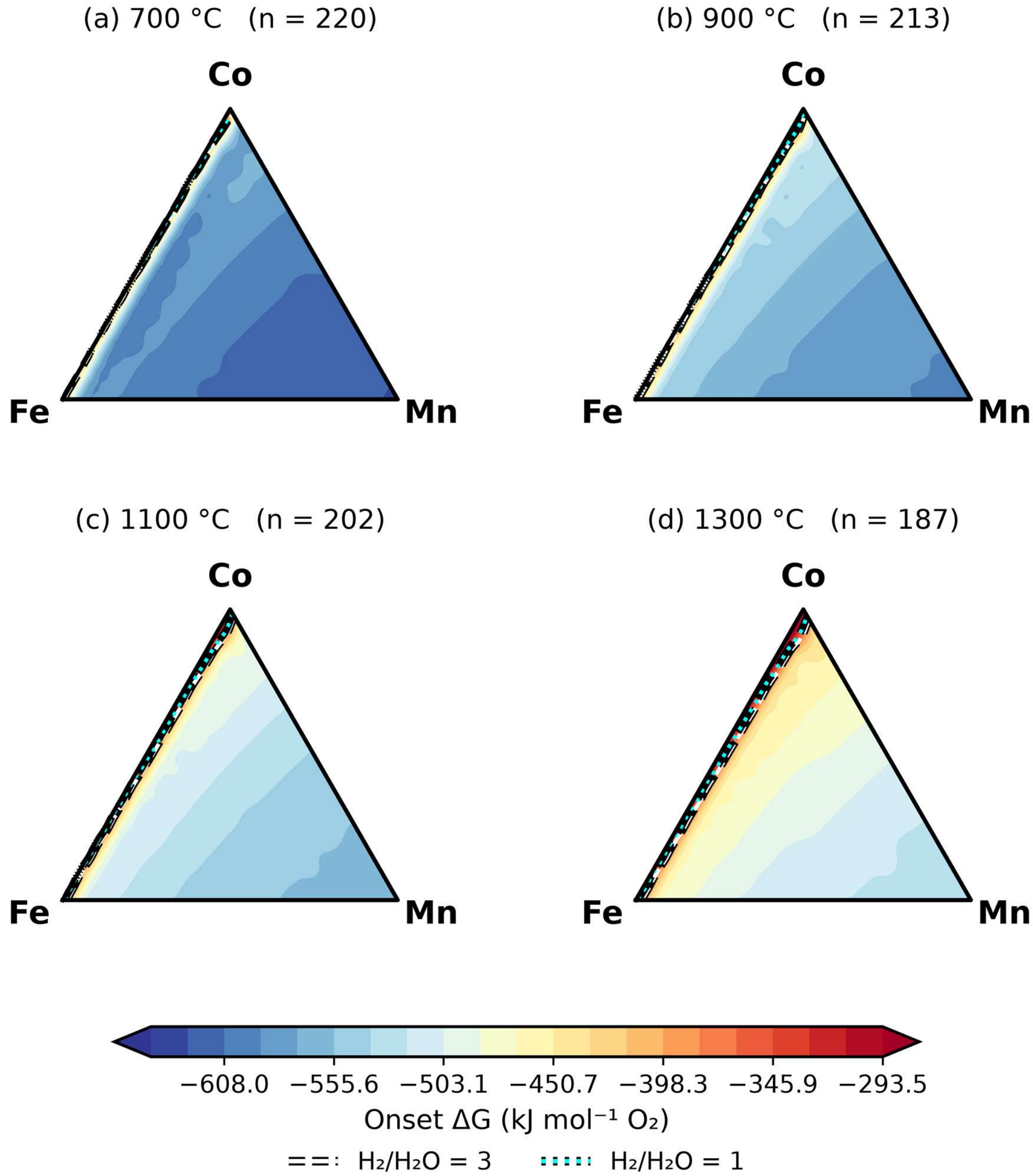

***Figure S5:*** *Onset ΔG on the Co–Fe–Mn face. Mn lowers the onset energy and limits the favourable region to 7–8% of sampled points at* $H_2/H_2O$ *= 3. Here n counts stored onset calculation records at that temperature, including repeated compositions; see **Table S5** for unique-composition counts.*

## Oxide phase identity

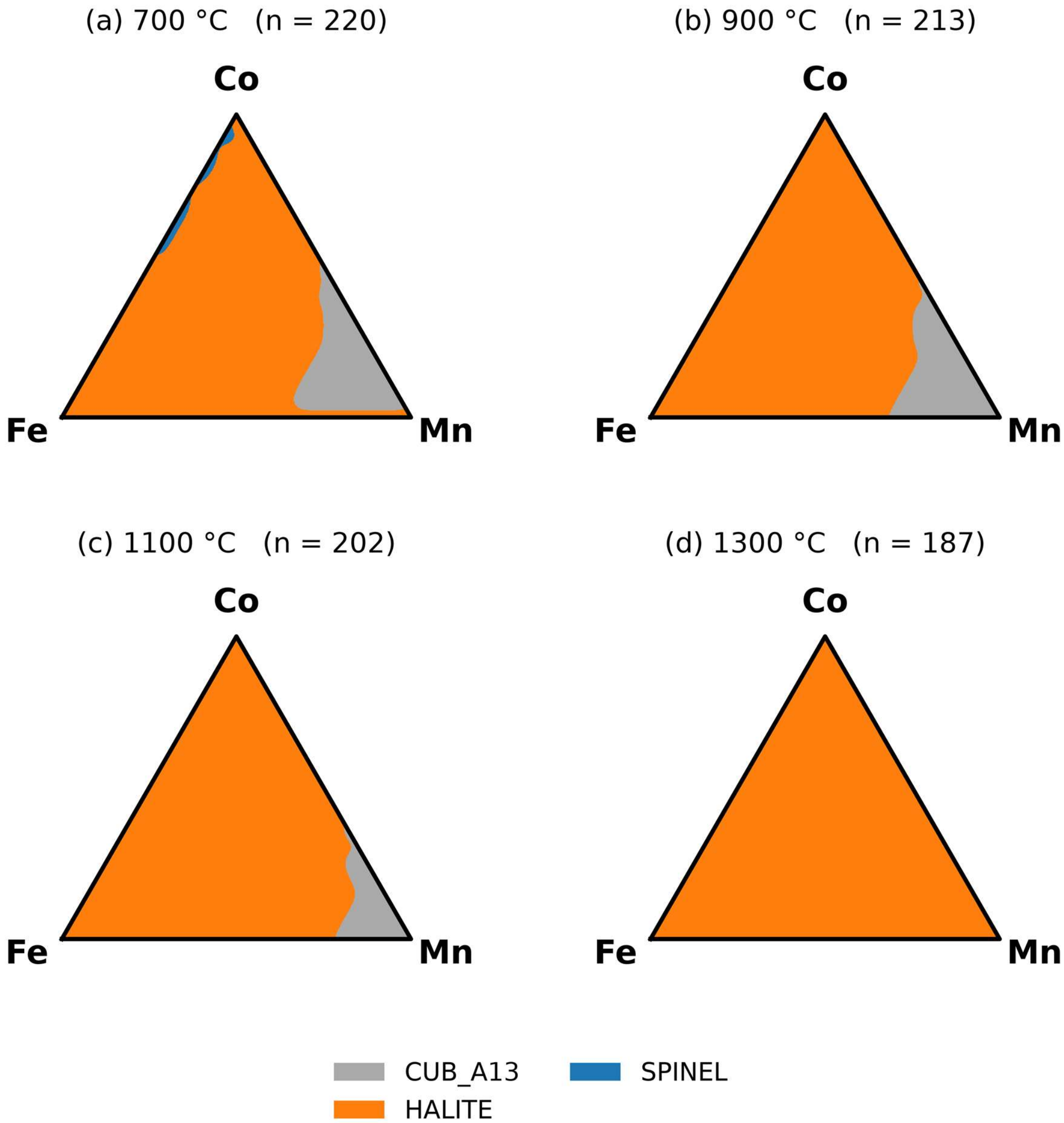

***Figure S6:*** *First oxide on the Co–Fe–Mn face. HALITE dominates, CUB_A13 occupies the Mn-rich region at lower temperature, and the small SPINEL field disappears with increasing temperature. Here n counts onset records passing the phase-label screen at that temperature, including repeated compositions; see* ***Table S5****.*

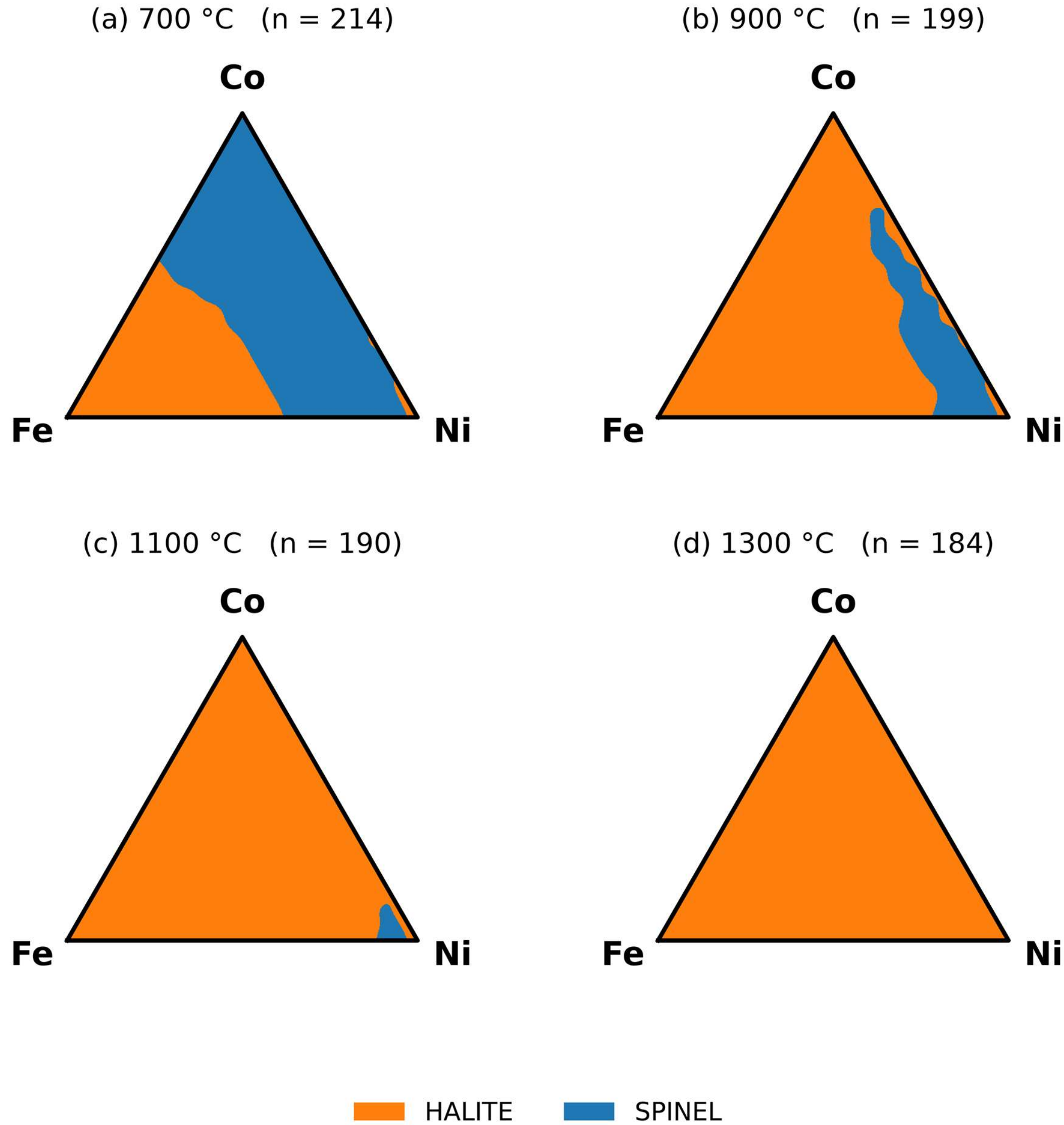

***Figure S7***: *First-listed oxide on the Co–Fe–Ni face. The mapped SPINEL region contracts as temperature increases and is not resolved in the 1300 °C panel. Here n counts onset records passing the phase-label screen, including repeated compositions; absence of a resolved colored field is not a test of zero phase amount. See* ***Table S5*** *for the corresponding populations.*

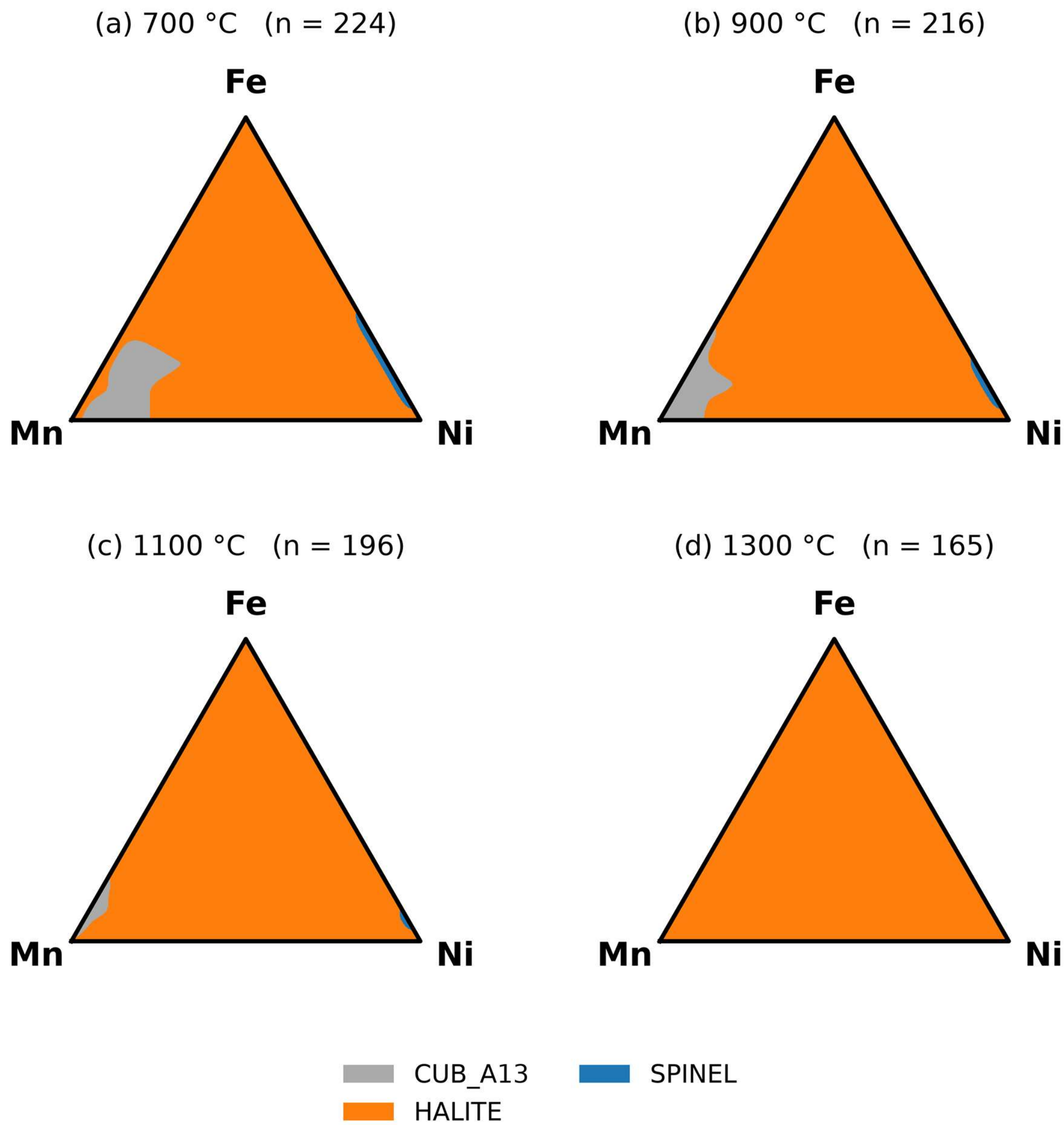

***Figure S8:*** *First oxide on the Fe–Mn–Ni face. HALITE covers most compositions; CUB_A13 and SPINEL form lower-temperature minority regions. Here n counts onset records passing the phase-label screen at that temperature, including repeated compositions; see* ***Table S5***.

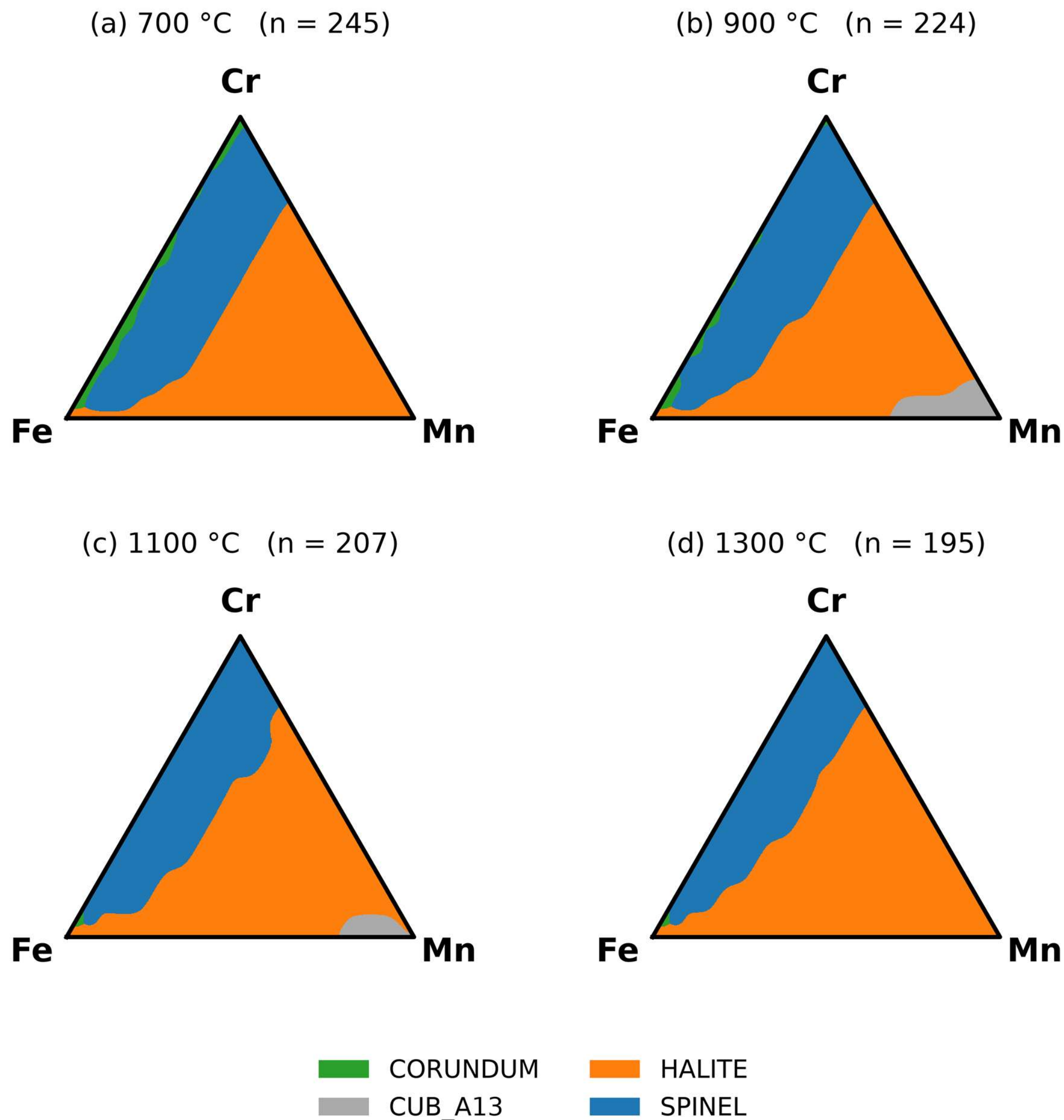

***Figure S9:*** *First oxide on the Cr–Fe–Mn face. HALITE, SPINEL, CORUNDUM, and a small CUB_A13 region partition the face. Here n counts onset records passing the phase-label screen at that temperature, including repeated compositions; see **Table S5**.*

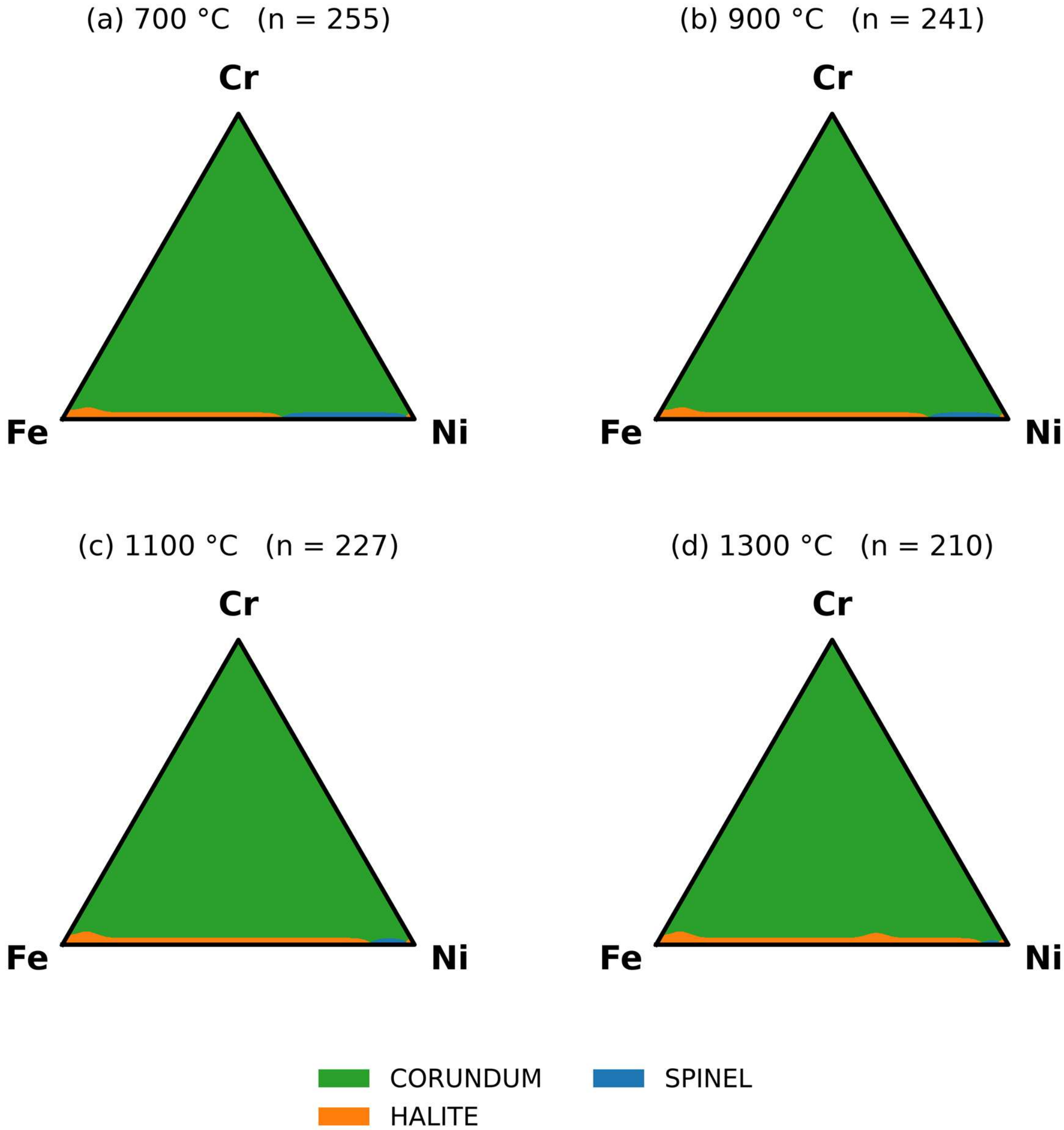

***Figure S10:*** *First oxide on the Cr–Fe–Ni face. CORUNDUM covers nearly the entire face at all four temperatures; HALITE and SPINEL occur only as narrow bands along the Fe–Ni edge, with HALITE toward Fe and SPINEL toward Ni. Here n counts onset records passing the phase-label screen at that temperature, including repeated compositions; see* ***Table S5****.*

## Regression coefficient table

***Table S3**: Standardized ordinary least-squares (OLS) coefficients for 35,934 valid Fe-bearing onset records. Ni is the reference composition variable. $\beta_k$ denotes the standardized coefficient of element k defined by Equation (9) of the main text and t the conventional coefficient-to-standard-error ratio under an independent-residual assumption. Because temperatures and repeated records share compositions, these t statistics are descriptive; **Table S8** gives composition-grouped uncertainty estimates.*

| | $\log_{10}(p_{O_2})$ | | $\Delta G_{onset}$ | |
|---|---|---|---|---|
| Predictor | $\beta_k$ | $t$ | $\beta_k$ | $t$ |
| Temperature | +0.928 | +520 | +0.667 | +216 |
| Mn | -0.200 | -98 | -0.459 | -130 |
| Cr | -0.169 | -76 | -0.395 | -102 |
| Fe | -0.060 | -27 | -0.146 | -38 |
| Co | -0.005 | -2 | -0.007 | -2 |
| $R^2$ | 0.887 | | 0.659 | |
| $n$ | 35,934 | | 35,934 | |

The pooled OLS model fits each response to standardized temperature and Co, Cr, Fe and Mn fractions plus an intercept; Ni is omitted because the metal fractions sum to unity. Both predictors and response are standardized using sample standard deviations. A coefficient therefore describes a conditional change relative to the omitted Ni fraction, not an unconstrained addition at fixed concentrations of every other metal. The rows contain deterministic calculations with repeated compositions, not independent experiments. Temperature-resolved and binary-host sensitivity plots in the manuscript complement this pooled summary.

***Table S4:** Nested composition-blocked performance of the nonlinear model. Here n denotes held-out calculation records, including repeated compositions, rather than independent experimental samples.*

| Subset | $n$ | $R^2$ | RMSE | MAE |
|---|---|---|---|---|
| | | | $kJ\ mol^{-1}\ O_2$ | |
| All higher-order | 19,817 | 0.987 | 6.81 | 2.16 |
| Quaternary | 15,351 | 0.984 | 7.62 | 2.25 |
| Quinary | 4,466 | 0.998 | 2.54 | 1.82 |

## Nonlinear model diagnostics

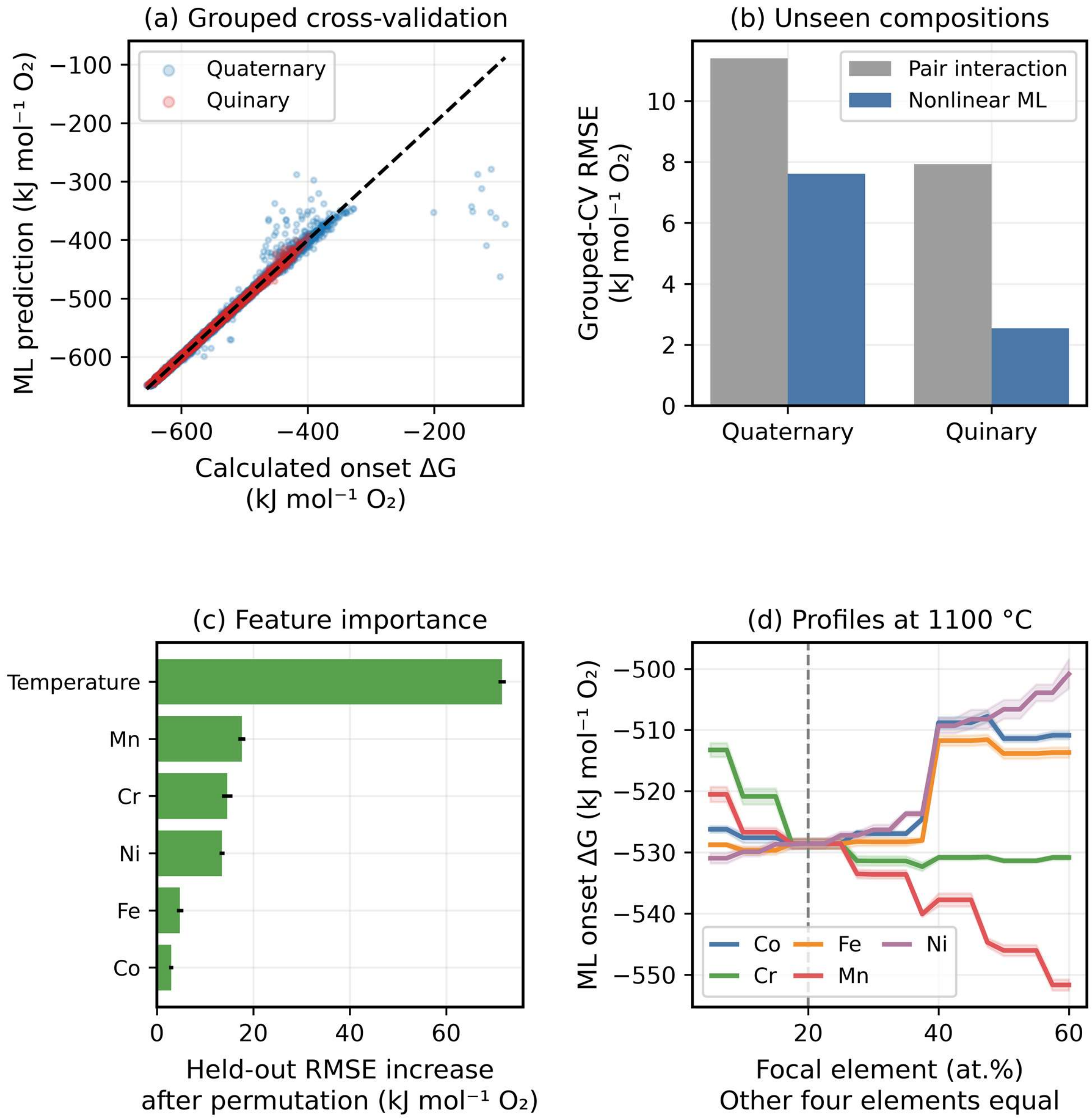

***Figure S11****: Statistical predictions of calculated onset energy. (a) Out-of-composition predictions from nested grouped cross-validation; the dashed line is equality. (b) RMSE comparison with the pair-interaction model, whose tuning and reporting use the same folds and therefore a different assessment protocol. (c) Closure-preserving permutation importance; error bars are one standard deviation across five outer folds and eight repeats per fold. (d) Composition profiles at 1100 °C; shading is one standard deviation across five fitted outer-fold models, not a prediction interval or database uncertainty. All panels describe computational data.*

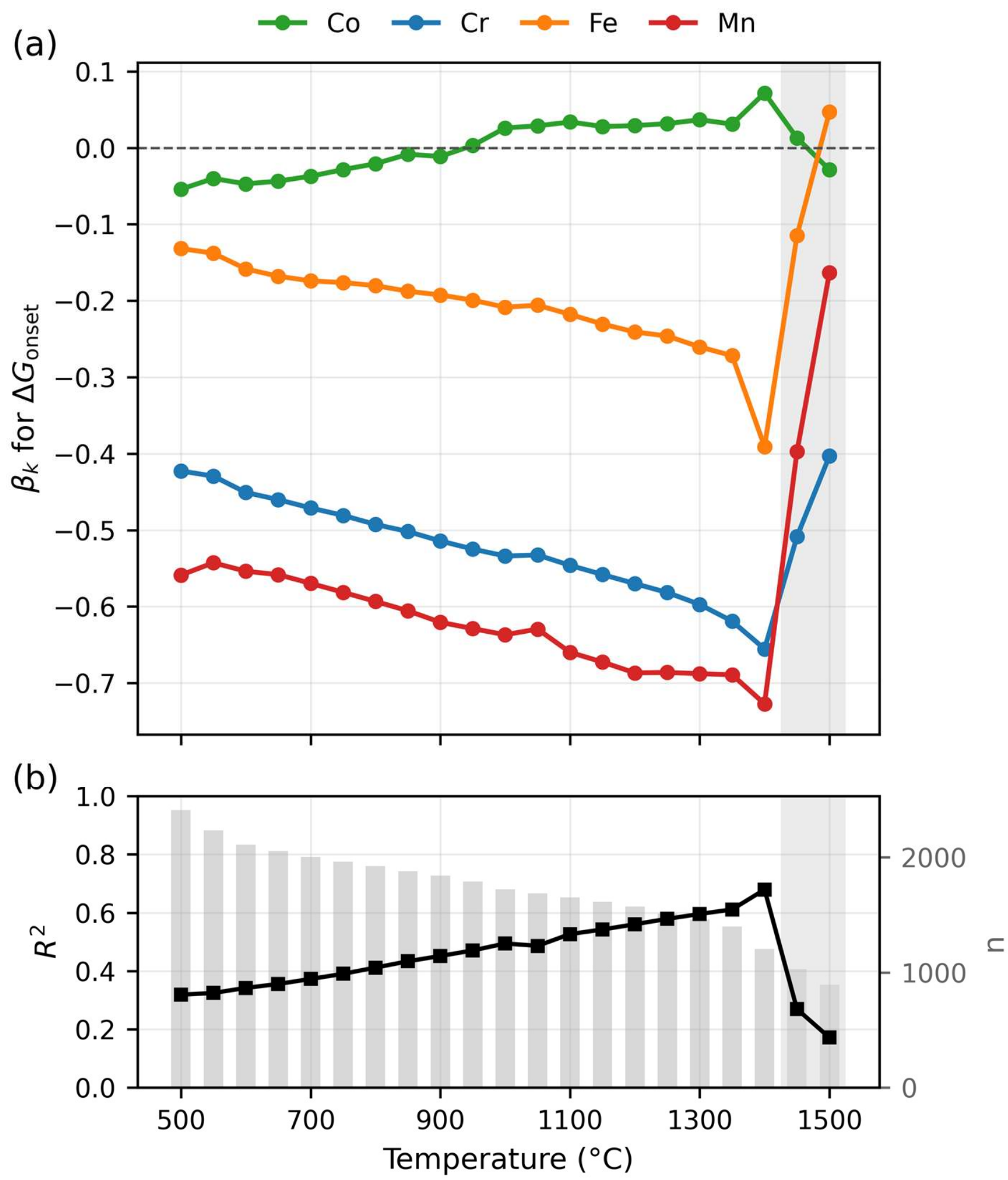

***Figure S12****: Temperature-resolved standardized coefficients for the valid Fe-bearing onset records. (a) $\beta_k$ for Co, Cr, Fe and Mn (Ni reference), obtained by fitting Equation (9) of the main text, without the temperature term, separately at each temperature; negative values denote oxide stabilization relative to Ni. (b) Coefficient of determination $R^2$ (line, left axis) and number of valid Fe-bearing onset records n (bars, right axis) in each temperature slice. Variables are standardized within each slice, so the coefficients compare relative effects at one temperature rather than absolute composition derivatives. The shaded 1450 and 1500 °C slices combine fewer records with low $R^2$ and are not interpreted.*

## Calculation dataset and coverage

The saved calculation dataset (checkpoint) contains 310,749 stored boundary records, but not that many independent calculations. Boundary index 1 identifies 55,051 onset records; the phase-label screen retains 54,385, of which 35,934 contain Fe. A valid label is a nonempty string for which each token separated by a vertical bar contains at least two characters. This is a parsing-quality screen, not a thermodynamic convergence test. Coexisting labels occur in 1399 valid onset records (2.6%); their first-listed phase is used for categorical summaries, so frequencies depend on this reporting convention. Phase-resolved analyses use valid labels; numerical onset maps and stored-boundary complexity summaries have their own populations and must not be interpreted as identically filtered datasets. The 3526-composition checkpoint is not a complete Cartesian product of compositions and temperatures.

To put these numbers into perspective, the dataset covers 3,526 metal compositions at 21 temperatures (500–1500 °C in 50 °C steps), i.e., 74,046 composition–temperature pairs. For each pair, the oxygen chemical-potential scan can detect several phase boundaries, each stored as one record, which is why the number of boundary records (310,749) far exceeds the number of pairs. Only the first boundary encountered from the metallic side (index 1, the onset) is used for the reduction thresholds. An onset was found for 52,078 pairs (70.3%); repeated calculations of the same pair add 2,973 duplicate onset records, giving the 55,051 onset records quoted above. For the remaining 21,968 pairs (29.7%), no onset was stored: 21,282 pairs returned only the no-boundary flag (index 0) and 686 pairs have no record. The missing fraction increases with temperature, from 305 of the 3,526 compositions at 500 °C to 1,957 at 1500 °C, and is slightly enriched in Co-rich compositions (mean Co content of 25 at.% for missing pairs versus 17 at.% for pairs with an onset). All statistics and maps in this work are therefore computed from the pairs with an onset, and the coverage per ternary face and temperature is reported in **Table S5**.

## Sampling audit for the supplementary maps

The following audit was calculated directly from the same checkpoint used to generate the SI maps. Every onset record has finite energy. For **Figures S1–S5**, n includes all onset records; for **Figures S6–S10**, n is the valid-label count. Distinct compositions are counted using the five metal percentages. At 1300 °C, one onset record is excluded from the phase maps on each of the Fe–Mn–Ni and Cr–Fe–Mn faces. No excluded phase label is imputed.

***Table S5****: Panel populations and positive-driving-force records at $H_2/H_2O$ = 3. The final column is a record-weighted percentage before interpolation; n and unique counts must not be interpreted as experimental sample sizes.*

| Face | T °C | Onset n | Unique comps. | Valid labels | Positive records | Positive % |
|---|---|---|---|---|---|---|
| Fe-Mn-Ni | 700 | 224 | 170 | 224 | 32 | 14.29 |
| Fe-Mn-Ni | 900 | 216 | 168 | 216 | 32 | 14.81 |
| Fe-Mn-Ni | 1100 | 196 | 166 | 196 | 29 | 14.80 |
| Fe-Mn-Ni | 1300 | 166 | 155 | 165 | 24 | 14.46 |
| Cr-Fe-Mn | 700 | 245 | 222 | 245 | 2 | 0.82 |
| Cr-Fe-Mn | 900 | 224 | 211 | 224 | 3 | 1.34 |
| Cr-Fe-Mn | 1100 | 207 | 204 | 207 | 1 | 0.48 |
| Cr-Fe-Mn | 1300 | 196 | 195 | 195 | 1 | 0.51 |
| Cr-Fe-Ni | 700 | 255 | 221 | 255 | 32 | 12.55 |
| Cr-Fe-Ni | 900 | 241 | 211 | 241 | 33 | 13.69 |
| Cr-Fe-Ni | 1100 | 227 | 204 | 227 | 29 | 12.78 |
| Cr-Fe-Ni | 1300 | 210 | 203 | 210 | 24 | 11.43 |
| Co-Fe-Ni | 700 | 214 | 198 | 214 | 214 | 100.00 |
| Co-Fe-Ni | 900 | 199 | 183 | 199 | 199 | 100.00 |
| Co-Fe-Ni | 1100 | 190 | 178 | 190 | 190 | 100.00 |
| Co-Fe-Ni | 1300 | 184 | 178 | 184 | 184 | 100.00 |
| Co-Fe-Mn | 700 | 220 | 198 | 220 | 18 | 8.18 |
| Co-Fe-Mn | 900 | 213 | 195 | 213 | 16 | 7.51 |
| Co-Fe-Mn | 1100 | 202 | 194 | 202 | 14 | 6.93 |
| Co-Fe-Mn | 1300 | 187 | 182 | 187 | 14 | 7.49 |

## Model specification and error diagnostics

The nonlinear model uses 19,817 valid onset records spanning 1,377 distinct quaternary or quinary compositions. Inputs are the five metal fractions and reduced temperature (T°C − 1000)/500. Five outer GroupKFold partitions hold all temperatures and repeated rows of each composition together; three inner grouped partitions choose hyperparameters using mean inner RMSE. GroupKFold is used without shuffling. The gradient-boosting estimator uses seed 20260731, squared-error loss, learning rate 0.06, 220 iterations, 128 bins and no early stopping.

The four tested (maximum leaves, minimum samples per leaf, L2 penalty) combinations are (15,30,1), (31,30,1), (15,60,3) and (31,60,5). **Table S6** reports the selected combination within each outer training set. These are reproducibility details of the stored analysis; no new model fitting was required for this audit. The saved out-of-fold observed responses and temperatures were checked against the current filtered checkpoint and agree row by row.

***Table S6****: Selected nonlinear-model hyperparameters. The inner RMSE guides selection within each outer training set; it is not the outer test error. RMSE units are kJ mol⁻¹ $O_2$.*

| Outer fold | Max. leaves | Min. samples | L2 penalty | Inner RMSE |
|---|---|---|---|---|
| 1 | 31 | 30 | 1 | 6.657 |
| 2 | 31 | 60 | 5 | 7.358 |
| 3 | 31 | 60 | 5 | 7.539 |
| 4 | 31 | 30 | 1 | 7.144 |
| 5 | 31 | 30 | 1 | 7.296 |

For the pair-interaction comparison, the inputs contain metal fractions, their products with reduced temperature and its square, and all ten pair products. Training-set centering and scaling precede ridge fitting. Composition groups are shuffled into five folds with seed 20260731; penalties 0, $10^{-4}$, $10^{-3}$, $10^{-2}$, 0.1, 1, 10 and 100 are compared, with 10 selected. Tuning and reporting use the same folds, unlike the nested nonlinear assessment. Their plotted errors are therefore informative comparisons, not a strictly matched benchmark.

***Table S7****: Absolute errors in saved nonlinear out-of-fold predictions, in kJ mol⁻¹ $O_2$. Percentiles are calculated over records in each subset, without removing outliers.*

| Subset | Median | 95th pct. | 99th pct. | Maximum |
|---|---|---|---|---|
| All higher-order | 1.42 | 5.35 | 9.40 | 367.06 |
| Quaternary | 1.44 | 5.46 | 10.27 | 367.06 |
| Quinary | 1.35 | 4.96 | 7.18 | 27.83 |

The overall RMSE of 6.81 kJ $mol^{-1}$ $O_2$ coexists with a maximum absolute error of 367.06 kJ $mol^{-1}$ $O_2$. The high overall $R^2$ does not establish reliable predictions at every composition. The error tail is retained and reported; no cause is assigned without further diagnosis. Predictions near a gas-threshold decision should be checked against direct equilibrium calculations, and extrapolation outside the sampled chemistry or temperature range is not established.

Permutation importance perturbs a focal element fraction using held-out values and rescales the remaining four fractions proportionally to preserve their sum. Temperature is permuted separately. The importance is the increase in held-out RMSE, averaged over folds and repeats. Profiles vary one metal from 5 to 60 at.% in 23 steps while sharing the remainder equally among the other four metals. Error bars and shading describe variation in this model procedure, not independent thermodynamic uncertainty.

## Composition grouped uncertainty in pooled regression

The original OLS point estimates in **Table S3** are retained. To assess dependence among records, a cluster sandwich covariance was recalculated with one group per unique metal composition, allowing arbitrary covariance among its temperatures and repeated rows. There are G = 2,209 groups, n = 35,934 records and k = 6 fitted coefficients including the intercept. The covariance uses the inverse predictor cross-product on each side of the sum of outer products of group-level residual score vectors, with finite-sample multiplier $[G/(G-1)] \times [(n-1)/(n-k)]$.

***Table S8****: Composition-grouped standard errors (SE) and coefficient/SE ratios for the* ***Table S3*** *standardized coefficients. These quantify uncertainty of a descriptive regression fit conditional on the sampled database surface; they are not uncertainty estimates for the thermodynamic database.*

| Predictor | $\log_{10}(pO_2)$ SE | $\log_{10}(pO_2)\beta_k$/SE | Onset ΔG SE | Onset $\Delta G\beta_k$/SE |
|---|---|---|---|---|
| Temperature | 0.0025 | 377.77 | 0.0040 | 166.26 |
| Mn | 0.0075 | -26.88 | 0.0162 | -28.37 |
| Cr | 0.0084 | -20.20 | 0.0186 | -21.23 |
| Fe | 0.0094 | -6.43 | 0.0205 | -7.10 |
| Co | 0.0098 | -0.50 | 0.0217 | -0.31 |

Grouping substantially reduces the magnitude of the coefficient/SE ratios relative to conventional independent-residual estimates. The negative Mn and Cr coefficients remain large relative to these grouped standard errors, whereas Co is small for both responses. This supports interpretation of the broad hierarchy but does not turn a deterministic composition grid into a random experimental sample. The estimates remain conditional on linear model form, sampled composition weighting and the specified metallic-phase model.

## Reproducibility and scope of the supplementary evidence

***Table S9***: *Files underlying the supplementary analyses. All data files are CSV files listed with their descriptions, row counts and checksums in data/catalog.json of the repository. The names identify calculation and analysis files available with the project; the manuscript data-availability statement governs access. This Word file contains the supporting summaries and does not embed the full calculation dataset.*

| Source file or script | Role |
|---|---|
| data/oxide_phase_boundaries.csv.gz | All stored phase-boundary records; metal percentages, temperature, boundary index, onset energy, oxygen pressure and phase labels. |
| plot_hydr_analysis.py; plot_hydr_extended.py | Continuous-map and categorical-map interpolation definitions. |
| analysis/nonlinear_ml_analysis.py; analysis/higher_order_sensitivity.py | Nonlinear nested assessment and pair-interaction comparison. |
| data/derived/nonlinear_model_heldout_predictions.csv; data/derived/nonlinear_model_nested_selection.csv | Stored predictions and inner selection scores; **Tables S4, S6 and S7**. |
| data/derived/nonlinear_model_permutation_importance.csv; data/derived/nonlinear_model_composition_profiles.csv | Importance means/standard deviations and composition profiles; **Figure S11**. |

The audit verifies finite onset energies, panel populations, phase-fit reconstruction, stored prediction alignment and stored error metrics. It does not provide an unrestricted metallic-phase comparison, a thermodynamic database-parameter uncertainty distribution, convergence against a uniformly finer composition grid, or proof that higher-index records form a monotonic reaction sequence. Repeated records remain in the stored model fits and record-weighted summaries; grouping prevents the same composition from appearing in training and test subsets, but does not give every composition equal weight.

The strongest quantitative use of the calculations is the composition-specific residual-oxide boundary under the stated equilibrium model. Smoothed fields, pooled phase lines, arithmetic boundary averages and surrogate scores are supporting descriptors. They cannot establish reduction times, phase amounts, transient ordering, precursor morphology or process emissions. Comparisons with published experiments are discussed in the manuscript and remain qualitative; the SI adds no proposed experimental campaign.